\documentclass[11pt]{article}
\usepackage{amsmath,amssymb,graphicx,youngtab,bm}
\usepackage{hyperref}
\usepackage{cite}

\usepackage{mathtools}

\usepackage{multicol,color,longtable}
\definecolor{darkred}{rgb}{0.65,0.15,0}
\hypersetup{pdfborder={0 0 0},colorlinks=true,urlcolor=blue,citecolor=blue,linkcolor=darkred,linktocpage=true}

\usepackage{array}
\newcolumntype{P}[1]{>{\centering\arraybackslash}p{#1}}
\usepackage{lscape,bbold}
\usepackage{multirow}
\usepackage{stmaryrd} 

\newcommand{\nn}{\nonumber}
\newcommand{\lb}{\left[}
\newcommand{\rb}{\right]}
\newcommand{\lc}{\left\{}
\newcommand{\rc}{\right\}}
\newcommand{\la}{\left\langle}
\newcommand{\ra}{\right\rangle}
\newcommand{\ci}{\text{i}\,}
\newcommand{\cK}{\mathcal{K}}
\renewcommand{\k}{\kappa}

\newcommand{\hI}{{\widehat{I}}}
\newcommand{\hJ}{{\widehat{J}}}
\newcommand{\hK}{{\widehat{K}}}
\newcommand{\hf}{\widehat{f}\,}
\newcommand{\hd}{\widehat{d}\,}

\newcommand{\CPoin}{\mathrm{cPoin}}
\newcommand{\adj}{\mathrm{adj}}
\newcommand{\X}{\mathbb{X}}
\newcommand{\Pp}{\mathbb{P}}
\newcommand{\Ad}{\mathrm{Ad}}
\newcommand{\tr}{\mathrm{tr}}
\newcommand{\Tr}{\mathrm{Tr}}

\newcommand{\id}{\mathbf{I}}

\newcommand{\reals}{\mathbb{R}}
\newcommand{\cx}{\mathbb{C}}
\renewcommand{\Im}{\mathrm{Im}}
\renewcommand{\Re}{\mathrm{Re}}

\newcommand{\mf}[1]{{\mathfrak{#1}}}

\newcommand{\eprint}[1]{{\href{http://arxiv.org/abs/#1}{[\texttt{#1}]}}}
\newcommand{\eprintN}[1]{{\href{http://arxiv.org/abs/#1}{[\texttt{#1 [hep-th]}]}}}

\newcommand{\doi}[2]{{\hypersetup{urlcolor=darkred}\href{http://dx.doi.org/#2}{#1}\hypersetup{urlcolor=blue}}}

\newcommand{\be}{\begin{equation}}
\newcommand{\ee}{\end{equation}}
\newcommand{\ba}{\begin{eqnarray}}
\newcommand{\ea}{\end{eqnarray}}
\newcommand{\eq}{&\!=\!&}

\def\cA{{\cal A}}
\def\cB{{\cal B}}
\def\cC{{\cal C}}

\def\cK{{\cal K}}

\def\cM{{\cal M}}

\def\cO{{\cal O}}
\def\cP{{\cal P}}

\makeatletter

\@addtoreset{equation}{section}
\makeatother

\begin{document}

 {\flushright  ICCUB-21-005\\
 Imperial-TP-KM-2021-01\\[20mm]
 }

\begin{center}

{\LARGE \bf \sc  Colourful Poincar\'e symmetry,\\[2mm] gravity and particle actions}\\[5mm]

\vspace{6mm}
\normalsize
{\large  Joaquim Gomis${}^{1}$, Euihun Joung${}^2$, Axel Kleinschmidt${}^{3,4}$ and Karapet Mkrtchyan${}^5$}

\vspace{10mm}
${}^1${\it Departament de F\'isica Qu\`antica i Astrof\'isica\\ and
Institut de Ci\`encies del Cosmos (ICCUB), Universitat de Barcelona\\ Mart\'i i Franqu\`es , ES-08028 Barcelona, Spain}
\vskip 1 em
${}^2${\it Department of Physics, Kyung Hee University\\Seoul 02447, Korea}
\vskip 1 em
${}^3${\it Max-Planck-Institut f\"{u}r Gravitationsphysik (Albert-Einstein-Institut)\\
Am M\"{u}hlenberg 1, DE-14476 Potsdam, Germany}
\vskip 1 em
${}^4${\it International Solvay Institutes\\
ULB-Campus Plaine CP231, BE-1050 Brussels, Belgium}
\vskip 1 em
${}^5${\it Theoretical Physics Group, Blackett Laboratory,\\ Imperial College London SW7 2AZ, U.K}
\vspace{20mm}

\hrule

\vspace{5mm}

 \begin{tabular}{p{14cm}}

We construct a generalisation of the three-dimensional Poincar\'e algebra that also includes a colour symmetry factor. This algebra can be used to define coloured Poincar\'e gravity in three space-time dimensions as well as to study generalisations of massive and massless free particle models. We present various such generalised particle models that differ in which orbits of the coloured Poincar\'e symmetry are described. Our approach can be seen as a stepping stone towards the description of particles interacting with a non-abelian background field or as a starting point for a worldline formulation of an associated quantum field theory.
\end{tabular}

\vspace{6mm}
\hrule
\end{center}

\thispagestyle{empty}

\newpage

\setcounter{page}{1}
\setcounter{tocdepth}{2}
\tableofcontents

\vspace{4mm}
\hrule

\section{Introduction}

Space-time symmetries that are larger than those realised in conventional gravitational systems,
including bosonic generators in non-trivial representations of isometry algebra are usually ruled out in field theories of finitely many interacting particles. The Coleman--Mandula theorem establishes that such symmetries are not compatible with a non-trivial S-matrix~\cite{Coleman:1967ad}. However, there are ways to evade the Coleman--Mandula theorem.

Einstein gravity in three dimensions can be written in Chern--Simons form \cite{Achucarro:1987vz,Witten:1988hc}. This formulation is based on gauging the global symmetry algebra and is background independent as well as directly related to the Einstein--Hilbert formulation of gravity,
if the dreibein is non degenerate.
Such a formulation can be generalised to higher-spin gravities without matter \cite{Blencowe:1988gj,Campoleoni:2010zq,Grigoriev:2020lzu} thus providing consistent examples of gravitational theories with extended symmetries. These theories lack bulk propagating degrees of freedom and in this way avoid no-go theorems based on S-matrix considerations. Other possible extensions that have been considered are given by Chern--Simons actions based on relativistic or non-relativistic Maxwell algebra extensions of the Poincar\'e algebra, see for example~\cite{Salgado:2014jka,Hoseinzadeh:2014bla,Papageorgiou:2010ud,Bergshoeff:2016lwr,Hartong:2016yrf,Aviles:2018jzw}.

A different extension of three-dimensional space-time symmetries is realised in \textit{coloured gravity} \cite{Gwak:2015vfb}. There, instead of adding generators corresponding to massless fields of spin $s\geq 3$, one has a colour extension of the isometry itself, with multiple copies of generators which correspond to Killing vectors and scalars associated to massless spin-two and spin-one fields. 
Such an extension is non-trivial for associative algebras of isometries \cite{Gwak:2015vfb,Gwak:2015jdo}, the first example was provided in \cite{Konstein:1989ij}. For the Poincar\'e algebra in arbitrary dimensions, such a colouring requires the colour algebra to be commutative and associative~\cite{Wald:1986bj,Cutler:1986dv,Wald:1986dw}, and together with the requirement of positive-definite bilinear form for the colour algebra (needed for unitarity) is trivial as shown in~\cite{Boulanger:2000rq}: the corresponding multi-gravity is described by a sum of mutually non-interacting Einstein--Hilbert actions.\footnote{In three space-time dimensions one can also deform the Poincar\'e algebra which gives a theory different from Einstein gravity, see~\cite{Boulanger:2000ni} where also an extension to a collection of spin-two fields governed by this deformed algebra was studied.} 
In three dimensions, the (A)dS${}_3$ isometry algebra, being a real form of $\mathfrak{sl}_2\oplus \mathfrak{sl}_2$, can be extended to an associative algebra by a double central extension. Incidentally, the same central extension allows taking different non-relativistic limits in three dimensions \cite{Joung:2018frr}.

It is a generic problem to couple matter to the gravitational systems with extended symmetries such as the coloured (higher-spin) gravity given by a Chern--Simons action in three dimensions.  Lagrangian formulations for known examples of higher-spin gravities with matter in three dimensions \cite{Prokushkin:1998bq,Prokushkin:1998vn} are not available yet\footnote{In \cite{Bonezzi:2015igv} an interesting proposal was made using a higher-dimensional space. See also \cite{Kazinski:2005eb} for a similar idea.}, despite some steps in that direction (see \cite{Fredenhagen:2019hvb} and references therein). In this paper we take a step back and instead try to  
understand it at the level of the worldline formulation of particles,
since it is the first step towards an associated quantum  field theory from a worldline approach.

We first note that in the theories with extended space-time symmetries the very notion of particle has to be reconsidered. Since particles in field theories as we know them are defined as (unitary) irreducible representations of the isometry algebra, for extended space-time symmetry algebras, one should consider the irreducible representations of the extended algebra. 
In principle, the representations of the larger symmetry algebra should decompose in terms of the representations of the original isometry algebra. We just note here, that in case that the representation of the larger algebra has higher Gelfand--Kirillov (GK) dimension
(see e.g. \cite{Vogan1978}
for the definition)
than that of regular particles, which will be true for the cases of our interest, the corresponding spectrum should be expected to contain infinite number of particles. 

A natural expectation is that systems with extended symmetries may allow for a mechanism of (spontaneous) symmetry breaking, that would lead to a regime with a more conventional gravitational system and massive particles coupled to it, much like in string theory. If such a scenario is realistic, then the notion of space and time is emergent and makes sense only for low-energy systems while at high energies (beyond the Planck scale?) it has to be abandoned or replaced by a more general notion. On the other hand, the higher symmetries of gravitational systems and their representations can be studied independently of the possible space-time interpretations.

We will show in this paper that not only the (A)dS${}_3$ but also the Poincar\'e algebra in three dimensions can be extended via a colour decoration. 
This example of the coloured Poincar\'e symmetry is interesting for several reasons.

In Minkowski space
the motion of a relativistic particle in a fixed electro-magnetic field  is described by the Lorentz force. In the case of constant electro-magnetic field the symmetry algebra is given by the Bacry--Combe--Richards algebra \cite{Bacry:1970ye}  that contains four space-time translations that do not commute, two Lorentz boost transformations and two central charges. Another possibility is to enlarge
the Poincar\'e algebra with
tensorial non-central charges, leading to the so-called Maxwell algebra \cite{Schrader:1972zd}. If one also enlarges Minkowski space with tensorial bosonic coordinates, the particle moving in a constant electro-magnetic field can be made invariant under the Maxwell symmetry
\cite{Bonanos:2008ez}. In an even further enlarged space-time one can generalise the Maxwell algebra to a free Maxwell algebra
\cite{Gomis:2017cmt} that can describe the  motion of the particle in a general electromagnetic field.

The motion of a relativistic particle with colour-flavour indices in a fixed Yang--Mills background, with space-time coordinates $x^a$ and colour coordinates $y_I$ (in the adjoint of $SU(N)$),
was first studied by Wong
\cite{Wong:1970fu}.
As in the electro-magnetic case it will be interesting to see 
whether the motion of a coloured particle in a fixed Yang--Mills background
has additional invariances associated with the colour coordinates, in particular, whether there exists  a non-abelian generalisation of the Maxwell algebra for the case of covariantly constant Yang--Mills background~\cite{Brown:1975bc,Batalin:1976uv}.

In this paper we take some first steps in this direction by constructing the coloured $(\mf{s})\mf{u}(N)$ Poincar\'e
algebra in three dimensions.\footnote{We comment on generalisations to higher dimensions in the conclusions.}
The generators are generalised commuting translations, internal symmetry transformations and coloured Lorentz transformations.
There is a natural generalisation of Minkowski space to a coloured Minkowski space with coordinates
$(x^a, y_I, x^a_I)$ where $x^a$ are the ordinary space-time coordinates, $y_I$ are the colour internal coordinates and 
$x^a_I$ are
coloured space-time coordinates. There is a coloured generalisation of the Poincar\'e transformations in this space. 
Since the space-time is enlarged, the usual Coleman--Mandula theorem no longer applies directly. 
Interestingly, one can restrict to the subspace defined by $x^a_I=0$ at the expense of breaking the full coloured Poincar\'e symmetry to Poincar\'e
and colour transformations. This subspace has the same coordinates as the Wong particle. 

In the coloured Minkowski space
we construct massive and massless  particle
actions that are invariant
under the coloured generalisation of the Poincar\'e algebra using a variety of methods. The methods differ in which coadjoint orbits of the coloured Poincar\'e algebra
we consider and we discuss their classification in some detail.\footnote{For a discussion of coadjoint orbits in the uncoloured case see for example \cite{Sou}.} The choice of orbit is at the heart of Wigner's construction of induced representations and presents the starting point for a particle interpretation.
If we restrict our constructions to the
subspace $x^a_I=0$ we can recover the free Wong equations.

We also construct the  particle action with a coloured AdS$_3$ symmetry \cite{Gwak:2015vfb,Gwak:2015jdo}, in the flat limit we recover the  coloured Poincar\'e particle found before. We also notice that 
subalgebra $\mf{u}(N,N)$ of coloured Poincar\'e, in three dimensions, with an appropriate identification of the  generators of coloured two-dimensional translations gives the coloured AdS$_2$ algebra. 
This algebra could be used to construct
the coloured AdS$_2$ particle or 
the one-dimensional coloured conformal particle mechanics that generalises to the coloured case the one of reference~\cite{deAlfaro:1976vlx}.

The organisation of the paper is as follows. We first discuss the algebraic method for colouring the three-dimensional Poincar\'e algebra in section~\ref{sec:CP3}. In section~\ref{sec:CSgrav}, we show that there is an invariant bilinear form on the $N$-coloured Poincar\'e algebra $\CPoin_3(N)$ and use it to construct a Chern--Simons theory corresponding to coloured Poincar\'e gravity. Section~\ref{sec:particle} discusses particle actions built from $\CPoin_3(N)$ that utilise coloured Minkowski that is also introduced there. Various types of particle actions differ by the co-adjoint orbits of $\CPoin_3(N)$ they describe. We also consider the coloured particle in an AdS background in section~\ref{sec:AdSpart} before offering some concluding remarks in section~\ref{sec:concl}. Several appendices contain complementary details.

\section{\texorpdfstring{Colouring the Poincar\'e algebra in $3$ space-time dimensions}{Colouring the Poincar\'e algebra in 3 space-time dimensions}}
\label{sec:CP3}

At the kinematic level, we first address the problem of adding colour indices to the Poincar\'e algebra, such that the new generators form a Lie algebra. Since the tensor product of two Lie algebras is not a Lie algebra, we have to resort to a different construction. In $D=3$ space-time dimensions a similar problem has been studied in the case of AdS and higher spin algebras~\cite{Gwak:2015vfb,Gwak:2015jdo} that we rely on. The construction is based on embedding the Poincar\'e Lie algebra into an associative algebra and then tensoring it with an associative `colour' algebra. Since the tensor products of associative algebras is associative and every associative algebra can be turned into a Lie algebra using the commutator, the construction yields a Lie algebra that contains a colouring of the original Poincar\'e Lie algebra. 

Before addressing the embedding the lifting of the Poincar\'e algebra to an associative algebra, we first discuss the colour algebra.
This we take to be $\mf{gl}(N,\mathbb{C})$
with $\mf{u}(N)$ as its basis over $\mathbb C$, corresponding to the subspace of all anti-hermitian $(N\times N)$ matrices.
The product in this basis is given by (see, e.g.,~\cite{Fuchs:1997jv})
\begin{align}
\label{eq:unass}
T^I \,T^J = -\frac{1}{N} \,\delta^{IJ}\, \id + \frac12\,(\ci d^{IJ}{}_K + f^{IJ}{}_K)\, T^K\,,\quad 
\id\,T^I = T^I \,\id = T^I\,,\quad \id \,\id =\id\,.
\end{align}
Here, $T^I$ are the traceless generators of $\mf{su}(N)$ with $I=1,\ldots, N^2-1$ and $\id$ is the $N\times N$ identity matrix. The structure constants $f^{IJ}{}_K$ in $\lb T^I,T^J\rb = f^{IJ}{}_K T^K$ are real while the real $d^{IJ}{}_K$ are the invariant tensors mapping the symmetric product of two $\mf{su}(N)$ adjoint representations back to the adjoint of $\mf{su}(N)$. They exist for $N>2$ and the identities satisfied by the invariant tensors turn $\mf{gl}(N,\mathbb{C})$ into an associative algebra. Here, $\delta^{IJ}$ is the invariant metric on the adjoint of $\mf{su}(N)$. When using this to raise indices, $d^{IJK}$ becomes totally symmetric and $f^{IJK}$ becomes totally antisymmetric. When colouring the Poincar\'e algebra later, we shall use only a real slice $\mf{u}(N)\subset \mf{gl}(N,\mathbb{C})$,
and for this reason we shall refer the associative algebra $\mf{gl}(N,\mathbb C)$ as $\mf{u}(N)$ with an abuse of notation.

To write the algebra in a uniform way, we denote the anti-hermitian element $T^0=\ci \id$ and use the indices $\hI=0,1,\ldots,N^2-1$ to label all the generators of $\mf{u}(N)$. The associative product is given by
\begin{align}
T^{\hI} \,T^{\hJ} = \frac12 \,\hf^{\hI\hJ}{}_{\hK} \,T^{\hK} + \frac{\ci}2 \,\hd^{\hI\hJ}{}_{\hK} \,T^{\hK} \,,
\end{align}
with real $\hf^{\hI\hJ}{}_{\hK}$ and $\hd^{\hI\hJ}{}_{\hK}$. The product decomposes into its anti-symmetric and symmetric parts according to
\begin{align}
\lb T^{\hI},  T^{\hJ}\rb  = \hf^{\hI\hJ}{}_{\hK} \,T^{\hK} \,,\quad\quad
\lc T^{\hI},  T^{\hJ}\rc  =  \ci \hd^{\hI\hJ}{}_{\hK}\, T^{\hK} \,.
\end{align}
The tensors $\hf^{\hI\hJ}{}_{\hK}$ and $\hd^{\hI\hJ}{}_{\hK}$ satisfy identities analogous to the usual $\mf{su}(N)$ identities. 
When splitting the adjoint indices they have components 
\begin{subequations}
\label{eq:dfsplit}
\begin{align}
\hf^{0\hI}{}_{\hK} &= 0\,, & \hf^{IJ}{}_0 &= 0 \,,& \hf^{IJ}{}_K &= f^{IJ}{}_K\,,\\
\hd^{0\hI}{}_{\hK} &= 2\,\delta^{\hI}_{\hK}\,, &\hd^{IJ}{}_0 &= \frac{2}{N}\,\delta^{IJ}\,,& \hd^{IJ}{}_K &= d^{IJ}{}_K\,.
\end{align}
\end{subequations}

The Poincar\'e Lie algebra in $(2+1)$ space-time dimensions has Lorentz and translations generators $M_a, P_a$ and commutation relations\footnote{\label{fn:eps}The indices $a,b,c=0,1,2$ are Minkowski space-time indices. We use $\varepsilon^{012}=+1$ and indices are raised and lowered with the $(-++)$ Minkowski metric.}
\begin{align}
\label{eq:Poin3}
\lb M_a, M_b \rb = \varepsilon_{ab}{}^c \,M_c \,,\quad\quad
\lb M_a, P_b \rb = \varepsilon_{ab}{}^c \,P_c \,,\quad\quad
\lb P_a, P_b \rb = 0 \,.
\end{align}
Its embedding into an associative algebra can be achieved as follows. The starting point is to write the generators of the Poincar\'e algebra as tensor products
\begin{align}
\label{eq:PoinTen}
M_a = L_a \otimes \mathcal{I} \,,\quad\quad 
P_a =  L_a \otimes \mathcal{J} \,,
\end{align}
where $\mathcal{I}$ and $\mathcal{J}$ are generators of an associative abelian algebra $\mf{a}$ (over $\reals$) satisfying
\begin{align}
\label{eq:IJ}
\mathcal{I}^2 =\mathcal{I} \,,\quad\quad
\mathcal{J}^2 =0\,,\quad\quad
\mathcal{I} \,\mathcal{J} = \mathcal{J}\, \mathcal{I} =\mathcal{J}\,,
\end{align}
while $L_a$ are generators of $\mf{sl}(2,\reals)\cong\mf{su}(1,1)$ satisfying $\lb L_a,L_b\rb = \varepsilon_{ab}{}^c\, L_c$.
For reality of the $\mf{sl}(2,\reals)$ structure constants $\varepsilon^{abc}$ we require $L_a^\dagger = -L_a$. For the anti-hermiticity of $M_a$ and $P_a$,  we assign to $\mathcal{I}$ and $\mathcal{J}$ the hermiticity properties\footnote{\label{fn:exp}The associative algebra~\eqref{eq:IJ} is the same as the semi-group $S_E^{(1)}=\{\lambda_0,\lambda_1,\lambda_2\}$ with $\lambda_i\cdot \lambda_j = \lambda_j\cdot \lambda_i=\lambda_{i+j}$ where $\lambda_j=0$ for $j>1$. One can use the Lie algebra expansion method~\cite{Hatsuda:2001pp,Boulanger:2002bt,deAzcarraga:2002xi,Izaurieta:2006zz} with this semi-group to obtain a new Lie algebra that is isomorphic to the Lie algebra we describe below. If we assign  the hermiticity properties $\lambda^\dagger_i=\lambda_i$, then the Lie algebra obtained by the expansion method is exactly the same as we present in~\eqref{eq:cPoin}.}
\begin{align}
\label{eq:IJherm}
\mathcal{I}^\dagger = \mathcal{I} \,,\quad \mathcal{J}^\dagger= \mathcal{J}\,.
\end{align}

Even though $\mf{sl}(2,\reals)$ is only a Lie algebra and not associative, the Lie bracket $[x\otimes a,y\otimes b] = \frac12\left( \lb x,y\rb \otimes \lc a,b\rc + \lc x,y \rc \otimes \lb a,b\rb\right) = [x,y]\otimes ab$ on $\mf{sl}(2,\reals)\otimes \mf{a}$ is well-defined since $\mf{a}$ is abelian,  a fact that was also used in the context of (conformal) gravity~\cite{Wald:1986dw,Boulanger:2002bt}.
The Lie algebra $\mf{sl}(2,\reals)\otimes \mf{a}$ can now be embedded in an associative algebra by extending the first factor and defining the eight-dimensional associative algebra
\begin{align}
\label{eq:AIJ}
\mf{A} = \mf{u}(1,1)\otimes \mf{a}
=\left\langle L_a\otimes \mathcal{I},\, L_a\otimes \mathcal{J},\, \ci \id_2 \otimes \mathcal{I}\,, \ci \id_2\otimes \mathcal{J}\right\rangle.
\end{align}
Here, the associative product on $\mf{u}(1,1)$ is given by
\begin{align}
L_a \,L_b = \frac12\, \varepsilon_{ab}{}^c \,L_c + \frac14 \,\eta_{ab} \,\id_2\,,
\end{align}
with $\id_2$ being the identity element of the algebra, with $\id_2^\dagger = \id_2$. This formula is similar to~\eqref{eq:unass}.

The coloured Poincar\'e algebra in $D=3$,  $\CPoin_3(N)$, is then defined to be the Lie algebra associated with the associative algebra given by the tensor product
\begin{align}
\label{eq:CPoin}
\CPoin_3(N) = \mf{A} \otimes \mf{u}(N)\,.
\end{align}
A basis of the vector space $\CPoin_3$ is given by the generators 
\begin{subequations}
\label{eq:cPoinbas}
\begin{align}
M_a^{\hI} & = -\ci L_a \otimes \mathcal{I} \otimes T^\hI\,,\\
P_a^{\hI} &=  -\ci L_a \otimes \mathcal{J} \otimes T^\hI\,, \\\
N^{\hI} &= \phantom{\ci}\id_2 \otimes \mathcal{I} \otimes T^\hI\,, \\
Q^{\hI} &=   \phantom{\ci}\id_2 \otimes \mathcal{J} \otimes T^\hI\,.
\end{align}
\end{subequations}
The factors of $\ci$ are introduced such that all generators are anti-hermitian.

The coloured Poincar\'e algebra then has the following commutation relations.
\begin{subequations}
\label{eq:cPoin}
\begin{align}
\lb M_a^{\hI}, M_b^{\hJ} \rb &= \frac{1}2 \,\varepsilon_{ab}{}^c \,\hd^{\hI\hJ}{}_{\hK} \,M_c^{\hK} - \frac14\, \eta_{ab}\, \hf^{\hI\hJ}{}_{\hK}\, N^{\hK}\,,\\
\lb M_a^{\hI}, N^{\hJ} \rb &= \hf^{\hI\hJ}{}_{\hK} \,M_a^{\hK}\,,\\
\label{eq:NN}
\lb N^{\hI}, N^{\hJ} \rb &= \hf^{\hI\hJ}{}_{\hK}\, N^{\hK}\,,\\
\lb M_a^{\hI}, P_b^{\hJ} \rb &= \frac{1}2 \,\varepsilon_{ab}{}^c \hd^{\hI\hJ}{}_{\hK}\, P_c^{\hK} -
 \frac14 \,\eta_{ab}\, \hf^{\hI\hJ}{}_{\hK}\, Q^{\hK}\,,\\
\lb M_a^{\hI}, Q^{\hJ} \rb &= \hf^{\hI\hJ}{}_{\hK} \,P_a^{\hK}\,,\\
\lb N^{\hI}, P_a^{\hJ} \rb &= \hf^{\hI\hJ}{}_{\hK} \,P_a^{\hK}\,,\\
\lb N^{\hI}, Q^{\hJ} \rb &= \hf^{\hI\hJ}{}_{\hK} \,Q^{\hK}\,,\\
\lb P_a^{\hI}, P_b^{\hJ} \rb &= 0\,,\\
\lb P_a^{\hI}, Q^{\hJ} \rb &= 0\,,\\
\lb Q^{\hI}, Q^{\hJ} \rb &= 0\,.
\end{align}
\end{subequations}
The first three lines are the algebra $\mf{u}(N,N)$ and the remaining lines can also be understood as the Lie algebra with the semi-group $S_E^{(1)}$ applied to $\mf{u}(N,N)$, see footnote~\ref{fn:exp}.

We see from the commutation relations~\eqref{eq:cPoin} that the $\hI=0$ components $M_a^0$ and $P_a^0$ satisfy the usual uncoloured Poincar\'e algebra and can be identified with these according to
\begin{align}
\label{eq:embPoin}
P_a \equiv P_a^0 \,,\quad M_a \equiv   M_a^0\,.
\end{align}

The Lie algebra $\CPoin_3(N)$ has a two-dimensional center that can be quotiented out. This center is spanned by $N^0$ and $Q^0$. From the definition~\eqref{eq:CPoin} we see that\footnote{We note that the result of the construction is the same as constructing the co-adjoint extension of $\mf{u}(N,N)$ using~\cite{Matulich:2019cdo,Barducci:2019jhj}.}
\begin{align}
\label{eq:CPiso}
\CPoin_3\cong \mf{u}(N,N)\inplus_{\adj} \mf{u}(N,N)\,,
\end{align}
where the notation indicates a semi-direct sum of the Lie algebra $\mf{u}(N,N)$ acting on another copy of the space $\mf{u}(N,N)$, now treated as an abelian algebra, via the adjoint action.\footnote{In this notation, the usual uncoloured Poincar\'e algebra is written as the semi-direct sum $\mf{so}(1,D-1)\inplus_{\mathrm{vec}} \reals^{1,D-1}$ in any space-time dimension $D$. The space $\reals^{1,D-1}$ is the abelian algebra of translations and acted upon by the Lorentz algebra using the $\mf{so}(1,D-1)$ vector representation.} The Lie algebra $\mf{u}(N,N)$ arises since $\mf{u}(1,1)\otimes \mf{u}(N)\cong \mf{u}(N,N)$ upon using $\mf{sl}(2,\reals)\cong \mf{su}(1,1)$. The non-abelian first $\mf{u}(N,N)$ in~\eqref{eq:CPiso} is due to the idempotent $\mathcal{I}$ in~\eqref{eq:AIJ} whereas the second abelian $\mf{u}(N,N)$ is due to the nilpotent $\mathcal{J}$. The two-dimensional center in this language corresponds to the two $\mf{u}(1)$ inside each of the two $\mf{u}(N,N)$,\footnote{The presence of two $\mf{u}(1)$
makes it possible to
contract $\CPoin_3(N)$ to
a non-relativistic colour algebra
with a non-degenerate bilinear form,
analogously 
to the uncolored case of
the extended Bargmann algebra
as a contraction of 
$\mf{u}(1,1)\inplus_{\adj} \mf{u}(1,1)$
\cite{Joung:2018frr}.
}
and we are left with the semi-direct sum
\begin{align}
\label{eq:sdnocent}
\mf{su}(N,N)\inplus_{\adj} \mf{su}(N,N)\,.
\end{align}

Because of the structure of~\eqref{eq:CPiso} we shall refer to the generators $M_a^\hI$ and $N^\hI$ as generalised (coloured) Lorentz generators and to $P_a^\hI$ and $Q^\hI$ as generalised (coloured) translation generators that form an abelian subalgebra and whose eigenvalues are generalised momenta. As is evident from~\eqref{eq:CPiso} and~\eqref{eq:sdnocent}, the coloured Poincar\'e algebra is not a direct sum of the usual Poincar\'e algebra $\mf{iso}(2,1)$ with an internal symmetry algebra. This is not in contradiction with the Coleman--Mandula theorem as we shall also enlarge the space-time beyond three dimensions when we construct particle actions below.

Since, according to~\eqref{eq:sdnocent}, all objects are elements of $\mf{su}(N,N)$, it will be convenient to use an explicit matrix representation. In the defining representation of $\mf{su}(N,N)$ an element of the Lie algebra $\mf{su}(N,N)$ is written in block matrix form as
\begin{align}
\label{eq:suNNmat}
\begin{pmatrix}
{\bf A} & {\bf B} \\ 
{\bf C} & {\bf D}
\end{pmatrix}
\end{align}
with $N\times N$ blocks satisfying
\begin{align}
{\bf A}^\dagger = - {\bf A}\,, \quad {\bf B}^\dagger ={\bf C} \,,\quad
{\bf C}^\dagger = {\bf B}\,, \quad {\bf D}^\dagger = -{\bf D}
\,,\quad \tr({\bf A} + {\bf D}) = 0
\,.
\label{2N rep}
\end{align}
We shall use bold letters to denote $(N\times N)$ matrices and `$\tr$' for the corresponding trace over $(N\times N)$-matrices. The identity $(N\times N)$ matrix is written as ${\bf I}$.

\subsubsection*{\texorpdfstring{Coloured AdS$_2$ algebra}{Coloured AdS2 algebra}}

The first three lines of~\eqref{eq:cPoin} are 
the $\mf{u}(N,N)$ subalgebra 
of $\CPoin_3(N)$, explicitly
\begin{subequations}
\label{eq:cAdS2}
\begin{align}
\lb M_a^{\hI}, M_b^{\hJ} \rb &= \frac{1}2 \,\varepsilon_{ab}{}^c \,\hd^{\hI\hJ}{}_{\hK} \,M_c^{\hK} - \frac14\, \eta_{ab}\, \hf^{\hI\hJ}{}_{\hK}\, N^{\hK}\,,\\
\lb M_a^{\hI}, N^{\hJ} \rb &= \hf^{\hI\hJ}{}_{\hK} \,M_a^{\hK}\,,\\
\lb N^{\hI}, N^{\hJ} \rb &= \hf^{\hI\hJ}{}_{\hK}\, N^{\hK}\,.
\end{align}
\end{subequations}
It
can be understood
as the tensor product
algebra $\mf{u}(1,1)\otimes \mf{u}(N)$\,:
\begin{subequations}
\label{eq:cAdSbas}
\begin{align}
M_a^{\hI} & = -\ci L_a \otimes T^\hI\,,\\\
N^{\hI} &= \phantom{\ci}\id_2 \otimes T^\hI\,.
\end{align}
\end{subequations}
With a proper identification of the generators, this algebra
is the coloured conformal algebra in one dimension or the coloured AdS$_2$ algebra.

The basis of the coloured conformal algebra is given by
\be 
\label{eq:Colconf}
D^{\hI}=M_0^{\hI}\,,
\qquad H^{\hI}=M_1^{\hI}+
M_2^{\hI}\,,\qquad 
K^{\hI}=M_1^{\hI}-
M_2^{\hI}\,.
\ee
The generators with the zero component in the colour indices give the uncoloured $D=1$ conformal algebra.

The coloured AdS$_2$ algebra can be written in terms of
conformal generators
\be 
P_1{}^{\hI}=\frac{1}{\ell}
D^{\hI}\,,
\qquad P_0{}^{\hI}=
\frac{1}{2\ell}(
H^{\hI}+
K^{\hI})\,,\qquad 
M_0{}^{\hI}=
\frac{1}{2}(
H^{\hI}-
K^{\hI})\,.
\ee
The parameter $\ell$ denotes the radius of AdS$_2$.
The generators with the zero component in the colour indices give the uncoloured AdS$_2$ algebra.
This algebra can be used in a BF theory \cite{Fukuyama:1985gg,Isler:1989hq,Chamseddine:1989wn,Alkalaev:2013fsa}
to construct a coloured 
analogue of Jackiw--Teitelboim gravity
(see the upcoming work \cite{AlkalaevJoungYoon}
for a detailed analysis).

\section{Coloured gravity in 3d Minkowski space-time}
\label{sec:CSgrav}

The coloured Poincar\'e algebra introduced in the previous section can be understood as a zero cosmological constant limit of the coloured (A)dS$_3$ algebra \cite{Gwak:2015vfb}, see also appendix~\ref{app:AdS}. Furthermore, one can define a Chern--Simons action for coloured gravity in 3d Minkowski space.

\subsection{Invariant bilinear form}

Let us recall the structure of the Poincar\'e algebra in three dimensions. As usual for Poincar\'e algebra in any dimensions, one can define a degenerate bilinear form
\ba
\la P_a\,|\,P_b\ra=0\,,\quad \la P_a\,|\,M_b\ra=0\,,\quad
\la M_a\,|\,M_b\ra=\eta_{ab}\,.
\ea
In three dimensions, there is a different, non-degenerate bilinear form (see, e.g., \cite{Witten:1988hc}) which is, however, non-diagonal
\ba
\la P_a\,|\,P_b\ra=0\,,\quad 
\la P_a\,|\,M_b\ra=\eta_{ab}\,,\quad
\la M_a\,|\,M_b\ra=0\,.\label{eq:PoinNDBF}
\ea
The latter bilinear form allows us to define the Chern--Simons action for Poincar\'e algebra, equivalent to the Einstein--Hilbert action in three dimensions for an invertible dreibein.
This is possible in three dimensions due to the fact that the Lorentz and translation generators of the Poincar\'e algebra carry the same Lorentz representation --- they are both vectors of $\mf{so}(1,2)$, or, equivalently, are in a representation isomorphic to the adjoint representation of $\mf{su}(1,1) \simeq \mf{so}(1,2)$. This property generalises to the coloured Poincar\'e algebra, which has an invariant bilinear form with similar properties as we show below. 

First we note that the most general bilinear form for the three-dimensional Poincar\'e algebra has a one-parameter freedom up to normalisation (see, e.g., \cite{Aviles:2018jzw}). Therefore, the most general Chern--Simons action based on three-dimensional Poincar\'e algebra can be given in the form,
\begin{align}
    S=\frac{\kappa_1}{4\,\pi}\int \Tr(\, e\wedge (d\omega+\omega\wedge \omega)\,)+\frac{\kappa_2}{4\,\pi}\int \Tr(\omega\wedge d\omega + \frac23 \omega\wedge\omega\wedge\omega )\,,\label{GeneralPoinGR3d}
\end{align}
where we do not write the contractions of fiber Lorentz indices for simplicity.
The second term in this action is the Lorentz--Chern--Simons term, which, supplemented with the torsion constraint describes conformal gravity in three dimensions. Vanishing of the torsion is a consequence of the equations of motion of the action \eqref{GeneralPoinGR3d}. However, substituting the solution of the torsion constraint back into the action will change its physical content since the torsionless constraint is not a consequence of the $\omega$ equations alone for $\kappa_2\neq 0$.
This situation is analogous to theories with ``third-way consistency" \cite{Bergshoeff:2015zga}. If one nevertheless substitutes the solution of the torsion constraint back in the action, one gets so-called Topologically Massive Gravity (TMG) \cite{Deser:1981wh} with mass value $m\sim\kappa_1/\kappa_2$.

In order to construct an invariant bilinear form for the coloured Poincar\'e algebra, we first start from the coloured Lorentz subalgebra $\mf{u}(N,N)\simeq \mf{u}(1,1)\otimes \mf{u}(N)$. For the colour group, $\mf{u}(N)$, the bilinear form has the following form
\ba
\la N^I\,|\, N^J\ra = \alpha\, \delta^{IJ}\,,\quad \la N^I\,|\, N^0\ra=0\,,\quad \la N^0\,|\, N^0\ra=\beta\,,
\ea
which is non-degenerate when $\alpha$ and $\beta$ are arbitrary non-zero numbers. A general ansatz for the rest of the generators, $M_a^{\hat I}=(M_a^I\,,\, M_a)$, can be given in the form:
\ba
\la M_a^I\,|\,M_b^J\ra=\gamma\,\eta_{ab}\,\delta^{IJ}\,,\quad \la M_a\,|\,M_b\ra=\delta\,\eta_{ab}\,.
\ea
The constraint of invariance of the bilinear form implies 
\ba
 \gamma=\frac14\,\alpha\,, \quad \delta=\frac{N}4\,\alpha\,.
\ea
As expected, the invariant bilinear form of the algebra $\mf{u}(N,N)$ depends on two arbitrary parameters $\alpha\,,\,\beta$. One of these numbers parametrises the norm of the central element, which we can drop.
Then, the bilinear form of the reduced coloured Lorentz algebra $\mathfrak{su}(N,N)$ is fixed uniquely up to normalisation:
\be
\la\mathcal{M}^A|\mathcal{M}^B\ra=\rho\,\delta^{AB}\,,\qquad \mathcal{M}^A\in \mathfrak{su}(N,N)\,,
\ee
where $A,B=1,2,\dots,4\,N^2-1$ denote here adjoint indices of $\mathfrak{su}(N,N)$. The most general invariant bilinear form for the algebra $\mathfrak{su}(N,N)\inplus_{\adj} \mathfrak{su}(N,N)$ is given in the following form.
\ba
\la\mathcal{M}^A|\mathcal{M}^B\ra=\rho\,\delta^{AB}\,,\quad \la\mathcal{M}^A|\mathcal{P}^B\ra=\lambda\,\delta^{AB}\,,\quad \la\mathcal{P}^A|\mathcal{P}^B\ra=0\,,
\ea
where $\rho$ and $\lambda$ are arbitrary numbers. This bilinear form is non-degenerate for $\lambda\neq 0$.

In analogy with \eqref{GeneralPoinGR3d}, one can write a Chern--Simons action for coloured Poincar\'e gravity in three dimensions in the following schematic form.
\begin{align}
    S=\frac{\kappa}{4\,\pi}\int \Tr(E\wedge (d\Omega+\Omega\wedge\Omega))+\frac{\tilde{\kappa}}{4\,\pi}\int \Tr(\Omega\wedge\,d\Omega+\frac23\Omega\wedge\Omega\wedge\Omega)\,,\label{GenCPoinGR}
\end{align}
where the fields $E,\,\Omega$ constitute the coloured Poincar\'e connection,
\begin{align}
    \cA=E^A\,\cP_A+\Omega^A\,\cM_A\,.
\end{align}
Naively, solving the torsion constraint,
\begin{align}
    T^{A}=dE^{A}+\tfrac12\,f^A{}_{BC}(\Omega^B\wedge E^C+E^B\wedge\Omega^C)=0\,,
\end{align}
for $\Omega=\Omega(E)$ and plugging back into the action \eqref{GenCPoinGR}, one will get the coloured analogue of TMG.  This substitution, even if possible technically, is leading to a different physical theory which we will not study here.

The space-time parity transformation multiplies odd space-time forms by $-1$ and transforms the coloured Poincar\'e algebra generators as $(\cP_A,\cM_A)\to (-\cP_A,\cM_A)$.
Consequently it 
acts on the one-form fields as  $(E,\Omega)\to (E,-\Omega)$ while $(dE,d\Omega) \to (-dE,d\Omega)$.
Under this action the first term of \eqref{GenCPoinGR} is invariant, while the second term changes sign.

For constructing the theory that is even under space-time parity,
one makes use of the non-diagonal bilinear form ($\rho=0$ case), which can be given in the following form (we write only non-zero scalar products).
\ba
\la M_a^{I}\,|\, P_b^{J}\ra=\tfrac14\,\lambda\,\eta_{ab}\,\delta^{IJ}\,,\quad \la N^{I}\,|\,Q^{J} \ra = \lambda\,\delta^{IJ}\,,\quad \la M_a\,|\, P_b\ra=\tfrac{N}{4}\,\lambda\,\eta_{ab}\,.\label{eq:CPBF}
\ea
Using this bilinear form, one can write a Chern--Simons action with  coloured Poincar\'e symmetry, that will be equivalent to the flat limit of coloured gravity in (A)dS$_3$, studied in \cite{Gwak:2015vfb}. 
It is natural to choose $\lambda=\frac{4}{N}$ to recover $\la M_a\,|\, P_b\ra=\eta_{ab}$ from  \cite{Witten:1988hc} (the overall constant is incorporated in the Chern--Simons level).

\subsection{Coloured gravity action in 3d Minkowski space}

As discussed above, the action for coloured gravity in three-dimensional Minkowski space-time can be given in the Chern--Simons form with coloured Poincar\'e gauge symmetry, using the non-degenerate bilinear form. There is one choice of the bilinear form, \eqref{eq:CPBF}, that reproduces \eqref{eq:PoinNDBF} for the uncoloured Poincar\'e subalgebra. We will use \eqref{eq:CPBF} to reproduce Einstein--Hilbert gravity for the $\mathfrak{su}(N)$ singlet spin-two field.

Coloured Poincar\'e gravity can be defined by a Chern--Simons action,
\begin{align}
    S[\mathcal{A}]=\frac{\kappa}{4\,\pi}\int \Tr\left(\mathcal{A}\wedge d\mathcal{A}+\frac23\,\mathcal{A}\wedge \mathcal{A}\wedge\mathcal{A}\right)\,,
\end{align}
where $\mathcal{A}\in \mf{su}(N,N)\inplus_{\adj}\mf{su}(N,N)$ (we factor by the center since it decouples in the action). In analogy with \cite{Gwak:2015vfb}, we will further decompose $\mathcal A$ as
\begin{align}
    \mathcal{A}=\mathcal{B}+\mathcal{C}\,,
\end{align}
where $\mathcal{B}$ incorporates the uncoloured isometry and colour group:\footnote{Here the one-form field $\omega$ is related to the standard spin-connection as $\omega^a=\frac12\epsilon^{abc}\omega_{bc}$. Similarly for its $\mathfrak{su}(N)$-valued analogue, $\omega^a_I$.}
\begin{align}
    \mathcal{B}=\omega^a\,M_a+e^a\,P_a+\varphi_I\,N^I+\tilde \varphi_I\,Q^I\,,
\end{align}
while $\mathcal{C}$ corresponds to the coloured spin-two sector:
\begin{align}
    \mathcal{C}=\omega^a_I\,M^I_a+e^a_I\, P^I_a\,.
\end{align}
The action then can be written in the form,
\begin{align}
    S[\mathcal{B},\mathcal{C}]=\frac{\kappa}{4\,\pi}\,\int \Tr\left(\mathcal{B}\wedge d\mathcal{B}+\frac23\,\mathcal{B}\wedge\mathcal{B}\wedge\mathcal{B}+\mathcal{C}\wedge D_{\mathcal{B}}\,\mathcal{C}+\frac23\, \mathcal{C}\wedge\mathcal{C}\wedge\mathcal{C}\right),
\end{align}
with
\begin{align}
    D_{\mathcal{B}}\,\mathcal{C}=d \mathcal{C}+\mathcal{B}\wedge\mathcal{C}+\mathcal{C}\wedge\mathcal{B}\,.
\end{align}
The sector that depends only on $\mathcal{B}$ can be explicitly written in the form,
\be
    S_{\mathcal{B}}=\frac{\kappa}{4\,\pi}\,\int \Tr\left(\mathcal{B}\wedge d\mathcal{B}+\frac23\,\mathcal{B}\wedge\mathcal{B}\wedge\mathcal{B}\right)=S_{\rm GR}+S_{\rm Vector}\,,
\ee 
with
\ba 
    S_{\rm GR}\eq   
    \frac{\kappa}{2\,\pi}\int e^a\wedge(d\omega^b+\tfrac12\,\epsilon^{bcd}\,\omega_c\wedge\omega_d)\,\eta_{ab}\,,\\
    S_{\rm Vector}\eq 
    \frac{2\,\kappa}{N\,\pi}\int \tilde \varphi^I\wedge(d\varphi^J+f^{J}{}_{KL}\,\varphi^K\wedge\varphi^L)\delta_{IJ}\,.
\ea
For zero cosmological constant, as opposed to the (A)dS$_3$ case, the rewriting  of the action in terms of chiral and anti-chiral pieces is not possible neither in the pure gravity sector, $S_{\rm GR}$, nor the vector field sector, $S_{\rm Vector}$. The equations of motion for the vector sector (neglecting contribution from interactions with coloured gravitons) can be given as
\begin{align}
    F(\varphi)=d\varphi+\varphi\wedge\varphi=0\,,\\
    D_{\varphi}\tilde\varphi=d
    \tilde\varphi+\varphi\wedge\tilde\varphi+\tilde\varphi\wedge\varphi=0\,,
\end{align}
and corresponds to a BF theory (see, e.g., \cite{Schwarz:1978cn,Horowitz:1989ng,Cattaneo:2000mc}).
For non-zero cosmological constant, $\Lambda\neq 0$, the Lagrangian of the GR sector would get an additional piece $\sim \Lambda\, e\wedge e\wedge e$, and the gauge vector field sector would get an additional piece $\sim \Lambda\,\tilde\varphi\wedge\tilde \varphi\wedge\tilde\varphi$, both essential for diagonalisation of the theory and deformation of the algebra.

The action of the matter sector has the following form,
\begin{align}
    S_{\mathcal{C}}=\frac{\kappa}{4\,\pi}\,\int \Tr\left(\mathcal{C}\wedge D_{\mathcal{B}}\,\mathcal{C}+\frac23\, \mathcal{C}\wedge\mathcal{C}\wedge\mathcal{C}\right)\,,
\end{align}
or, in a more explicit form,
\begin{align}
    S_{\cC}&=\frac{\kappa}{2\,N\,\pi}\int \Big(
    \delta^{IJ}\eta_{ab}\,e^a_{I}\wedge d\omega^b_J
    +\frac12\,\delta^{IJ}\,\epsilon_{abc}\,\omega^a_I\wedge\omega^b_J\wedge e^c \nn\\
    &\hspace{20mm}+ \delta^{IJ}\epsilon_{abc}\,\omega^a_I\wedge e^b_J\wedge \omega^c
    +\frac14 d^{IJK}\,\epsilon_{abc}\,\omega^a_I\wedge\omega^b_J\wedge e^c_K
    \nn\\
    &\hspace{20mm}- f^{IJK}\eta_{ab}\,\omega^a_I\wedge e^b_J\wedge \varphi_K
    - \frac12\,f^{IJK}\,\eta_{ab}\,\omega^a_I\wedge\omega^b_J\wedge\tilde{\varphi}_K
    \Big)\,.
\end{align}
It is interesting to note that the multiple background solutions with different cosmological constants of coloured gravity in (A)dS$_3$ \cite{Gwak:2015vfb} all go to the Minkowski solution in the $\Lambda\to 0$ limit.

The second term of \eqref{GenCPoinGR}, the coloured Lorentz--Chern--Simons term, is also straightforward to compute:
\begin{align}
    S_{\text{CLCS}}&=\frac{\tilde{\kappa}}{4\,\pi}\int (\omega^a\wedge\,d\,\omega^b\,\eta_{ab}+\tfrac13\,\epsilon_{abc}\,\omega^a\wedge\omega^b\wedge\omega^c)\nn\\
    &\quad+ 
    \frac{\tilde{\kappa}}{N\,\pi}\int (\varphi^I\wedge\,d\varphi^J\,\delta_{IJ}+\tfrac23\, f_{IJK}\,\varphi^I\,\wedge\varphi^J\wedge\varphi^K)\nn\\
    &\quad+\frac{\tilde\kappa}{4\,N\,\pi}\int \Big(
    \delta^{IJ}\eta_{ab}\,\omega^a_{I}\wedge d\omega^b_J
    +\tfrac16 d^{IJK}\,\epsilon_{abc}\,\omega^a_I\wedge\omega^b_J\wedge \omega^c_K\nn\\
    &\quad+\delta^{IJ}\,\epsilon_{abc}\,\omega^a_I\wedge\omega^b_J\wedge \omega^c 
    - f^{IJK}\eta_{ab}\,\omega^a_I\wedge \omega^b_J\wedge \varphi_K
    \Big)\,.
\end{align}

\section{Free coloured particle model}
\label{sec:particle}

In this section, we are considering massive and massless particle actions invariant under the coloured Poincar\'e algebra. First we introduce the coloured generalisation of Minkowski space and then discuss how the different ways of constructing Poincar\'e particle actions become coloured.  In our analysis we shall consistently quotient by the two-dimensional center generated by $N^0$ and $Q^0$
and thus restrict to $\mf{su}(N,N)\inplus_{\adj} \mf{su}(N,N)$ as given in~\eqref{eq:sdnocent}.

\subsection{Coloured Minkowski space}
\label{sec:colMink}

The algebra $\CPoin_3(N)$ is a vast extension of the Poincar\'e algebra~\eqref{eq:Poin3}. Its centerless version~\eqref{eq:sdnocent} includes an abelian part spanned by $P_a$, $P_a^I$ and $Q^I$. This should be thought of as coloured translations and there is an associated \textit{coloured Minkowski space}. Introducing dual coordinates for each of the coloured translation generators we obtain coordinates $x^a$, $x^a_I$ and $y_I$, where the first $x^a$ is to be thought of as the usual Minkowski coordinate (invariant under $\mf{su}(N)$ and the others are the coloured extensions. Similarly, the coloured Lorentz algebra $\mf{su}(N,N)$ has generators $M_a$, $M_a^I$ and $N^I$, where the first generator is to be thought of as the usual Lorentz generator and the others as coloured generalisations.

On a coloured Minkowski point $(x^a,x^a_I,y_I)$ represented by
\be 
\label{eq:ColMink}
	\mathbb X=x^a\,P_a+x^a_I\,P_a^I+y_I\,Q^I\,,
\ee
an infinitesimal coloured Poincar\'e transformation is obtained from the Lie algebra element
$(\mathbb O,\mathbb A)
\in \mf{su}(N,N)\inplus_{\adj} \mf{su}(N,N)$
with\footnote{The notation $\omega$ for the parameter here should not be confused with the gauge fields $\omega$ from section~\ref{sec:CSgrav}.}
\begin{align}
\label{eq:CPel}
	\mathbb O = \underbrace{\omega^a\, M_a + \omega^a_I \,M_a^I + \sigma_I\, N^I}_{\text{coloured Lorentz}}\,,
	\qquad \mathbb A=\underbrace{\alpha^a\, P_a + \alpha^a_I \,P_a^I+ \beta_I \,Q^I}_{\text{coloured translations}}\
\end{align}
and produces the linear transformation,
\be\label{colouredlorentz}
\delta \mathbb X=[\mathbb O, \mathbb X]+\mathbb A\,.
\ee
More explicitly, the transformation rule reads
\begin{subequations}
\label{eq:CM}
\begin{align}
\delta x^a &=  \varepsilon_{bc}{}^a \,\omega^b \,x^c +\frac{1}N \,\delta^{IJ} \,\varepsilon_{bc}{}^a\, \omega^b_I \,x^c_J +\alpha^a\,,\\
\delta x^a_I &=  \varepsilon_{bc}{}^a \,\omega^b x^c_I + \varepsilon_{bc}{}^a \,\omega^b_I\, x^c +\frac{1}2\,  \varepsilon_{bc}{}^a \,d^{JK}{}_I \,\omega^b_J\, x^c_K  +f^{JK}{}_I \,\omega^a_J \,y_K + f^{JK}{}_I\, \sigma_J \,x^a_K+\alpha^a_I\,,\\
\delta y_I &= -\frac14\, \eta_{ab}\, f^{JK}{}_I \,\omega^a_J \,x^b_K + f^{JK}{}_I \,\sigma_J y_K+\beta_I\,.
\end{align}
\end{subequations}
Here, we have used~\eqref{eq:cPoin}. 

The usual Lorentz part $\omega^a$ acts on the usual coordinate $x^a$ in the standard way. The coloured Lorentz generators $N^I$, that satisfy the $\mf{su}(N)$ algebra according to~\eqref{eq:NN}, generate colour rotations on the colour-extended coordinates $x^a_I$ and $y_I$. Note that these coordinates are in the adjoint of $\mf{su}(N)$.\footnote{It is amusing to observe that the transformation of the usual Minkowski coordinates $x^a$ does not depend on the remaining coordinates in the limit $N\to\infty$.}

The generators of coloured Lorentz group are obtained by tensoring such matrices with $\mathcal{I}$ and the generators of the coloured translations by tensoring with $\mathcal{J}$, see~\eqref{eq:IJ} and~\eqref{eq:cPoinbas}. We shall suppress the tensoring with these elements consistently. The action of the generalised Lorentz group on generalised Minkowski space can just be written as matrix operations. 

In the representation~\eqref{eq:suNNmat}, a point on coloured Minkowski space is written as
\begin{align}
\label{eq:Xmat}
\X = \begin{pmatrix}
\ci {\bf X}^0 + \ci {\bf Y} & {\bf X}^+ \\ 
{\bf X}^- & -\ci {\bf X}^0 + \ci {\bf Y}
\end{pmatrix}
\end{align}
with ${\bf X}^\pm = {\bf X}^1 \pm \ci {\bf X}^2$ and 
\begin{align}
\label{eq:Xpar}
{\bf X}^a &=  x^a \,{\bf I} + \ci x^a_I \,T^I\,, &
{\bf Y} &= \ci y_I \,T^I\,.
\end{align}
All coefficients are real-valued, thus $\left({\bf X}^\pm\right)^\dagger = {\bf X}^{\mp}$.
Writing the parameter of a transformation~\eqref{eq:CPel} as
\begin{align}
    \mathbb{O} = \begin{pmatrix}
    \ci \boldsymbol{\omega}^0 + \ci \boldsymbol{\sigma} & \boldsymbol{ \omega}^+ \\
    \boldsymbol{\omega}^- & -\ci \boldsymbol{\omega}^0 + \ci \boldsymbol{\sigma} 
    \end{pmatrix}\,,\qquad
    \mathbb{A} = \begin{pmatrix}
    \ci \boldsymbol{\alpha}^0 + \ci \boldsymbol{\beta} & \boldsymbol{ \alpha}^+ \\
    \boldsymbol{\alpha}^- & -\ci \boldsymbol{\alpha}^0 + \ci \boldsymbol{\beta}
    \end{pmatrix},
\end{align}
with component expansions similar to above, the transformation in~\eqref{colouredlorentz} corresponds exactly to the matrix commutator.

A metric on coloured Minkowski space that is invariant under coloured Lorentz transformations is given by
\begin{align}
\label{eq:invmetric}
\frac1{2N}\Tr (\X^2) 
= x^a x^b \eta_{ab}  +\frac{1}{N} x^a_I x^b_J \eta_{ab} \delta^{IJ} - \frac{1}{N} y_I y_J \delta^{IJ}\,,
\end{align}
where $\eta_{ab}$ is the $(-++)$ Minkowski metric and $\delta^{IJ}$ the $SU(N)$ invariant metric.
We see that it consists of the usual Minkowski metric for the coordinates $x^a$ and the other coordinates are suppressed by factors of $1/N$.
Using the invariant metric~\eqref{eq:invmetric} one can define the notion of coloured null, time-like and space-like vectors in the usual way depending on whether the norm is zero, negative or positive, respectively.

For writing particle actions, we shall also make use of the matrix notation~\eqref{eq:Xmat} and moreover require a matrix representation of the (conjugate) momentum that we write as
\begin{align}
\label{eq:Pmat}
\mathbb{P} = \begin{pmatrix} \ci {\bf P}^0 +\ci {\bf \Pi} & {\bf P}^+ \\  {\bf P}^- & -\ci {\bf P}^0 +\ci {\bf \Pi} \end{pmatrix}\,.
\end{align}
The momentum vector $\mathbb{P}$ transforms in the coadjoint of the coloured Lorentz algebra $\mf{su}(N,N)$, thus $\mathbb{P}\in \mf{su}(N,N)^*$:
$\delta_{\mathbb{O}} \mathbb{P}=[\mathbb{O},\mathbb{P}]$.
The components satisfy $\left({\bf P}^\pm\right)^\dagger = {\bf P}^\mp$ and ${\bf \Pi}$ 
is traceless and hermitian. Expanding out these matrices explicitly we write
\begin{align}
\label{eq:Ppar}
{\bf P}^a = p^a\, \id + \ci p^a_I \,T^I\,,\quad
{\bf \Pi} = \ci \pi_I \,T^I\,.
\end{align}
We note that in the uncoloured case $N=1$ we have
\begin{align}
\label{eq:Pmat1}
    N=1:\hspace{10mm} \mathbb{P}^2 = \begin{pmatrix}
     p^ap_a & 0 \\ 
     0 & p^a p_a
    \end{pmatrix} = p^a p_a \mathbb{I}\,.
\end{align}
That $\mathbb{P}^2$ is proportional to the identity matrix is unique to the case $N=1$.

We shall moreover need a matrix Lagrange multiplier $\mathbb{L}$ whose matrix form we take as 
\begin{align}
\label{eq:Lmat}
\mathbb{L} = -\ci \begin{pmatrix} \ci {\bf K}^0 +\ci {\bf \Lambda} & {\bf K}^+ \\ {\bf K}^- & -\ci {\bf K}^0 +\ci {\bf \Lambda} \end{pmatrix}\,,
\end{align}
that satisfies $\left({\bf K}^\pm\right)^\dagger = {\bf K}^\mp$ and where $\bf \Lambda$ is now traceful and hermitian. In other words $\ci \mathbb{L}\in\mf{u}(N,N)$ 
and we have multiplied by a convenient $-\ci$ for simplifying certain expressions later. 
Expanding out the matrices we write
\begin{align}
\label{eq:Lpar}
{\bf K}^a = f^a\, {\bf I} + \ci k_I^a \,T^I\,,\quad
{\bf \Lambda} = e\, {\bf I } + \ci \lambda_I \,T^I \,.
\end{align}
The extra $\mf{u}(1)$ component corresponds to the parameter $e$.

\subsection{Different types of massive particle actions}

Before discussing particle actions for coloured Minkowski space we briefly review different particle actions in uncoloured Minkoswki space and how they are related.

\subsubsection{Uncoloured Minkowski space}

In ordinary Minkowski space with coordinates $x^a$, there are several ways of writing the action for a massive particle of mass $m>0$. 
\begin{enumerate}
    \item 
    \underline{Geometric action}: 
    \begin{align}
    \label{eq:Lgeouc}
      S_{\text{geo}}^{\text{unc.}}[x^a] =  -m \int d\tau \sqrt{-\dot{x}^a \dot{x}^b \eta_{ab}}\,.
    \end{align}
    This action describes the proper length of the time-like world-line ($\dot{x}^a \dot{x}^b \eta_{ab}<0$), where the dot denotes differentiation with respect to the arbitrary world-line parameter $\tau$. The action is invariant under general world-line reparametrisations. We note that the momentum conjugate to $x^a$ that follows from this action is
    \begin{align}
        p_a = \frac{m \dot{x}_a}{\sqrt{-\dot{x}^b \dot{x}^c \eta_{bc}}} 
        \hspace{10mm} \Longrightarrow\quad
        p_ap^a +m^2 = 0\,.
    \end{align}
    The last equation is the usual mass-shell constraint for a massive particle. In proper time gauge $\dot{x}^a \dot{x}_a =-1$, the equations of motion are simply $\ddot{x}^a=0$.
   \item 
   \underline{Hamiltonian action}: 
   \begin{align}
   \label{eq:LHamuc}
        S_{\text{Ham}}^{\text{unc.}}[x^a, p_a,e]=\int d\tau \left( \dot{x}^a p_a + e( p^a p_a + m^2)\right).
   \end{align}
   This action utilises the einbein $e$ and the (conjugate) momentum $p_a$, both of which appear algebraically and are fully independent variables. By first integrating out $p^a$ and then $e$ by using their equations of motion one re-obtains~\eqref{eq:Lgeouc}. Since the equation for $e$ is quadratic there are two solutions, corresponding to changing the overall sign of~\eqref{eq:Lgeouc}. The mass-shell constraint,
   \begin{align}
   \label{eq:shelluc}
       p^ap_a + m^2=0\,,
   \end{align}
   obtained by varying~\eqref{eq:LHamuc} with respect to $e$ has two independent solutions (up to the action of the orthochronous Lorentz group) corresponding to positive and negative energy particles that are distinguished by the sign of $p^0$. More abstractly, there are two independent semi-simple orbits of the action of $SO(2,1)$ on the space of momenta $p^a$ that satisfy the uncoloured mass-shell constraint~\eqref{eq:shelluc}.
   
   If we write the uncoloured Minkowski coordinates in matrix form using~\eqref{eq:Xmat}, we can also write this action as
   \begin{align}
   \label{eq:canuc}
       S_{\text{Ham}}^{\text{unc.}}[ \X,\Pp,e] = \frac12 \int d\tau \Tr \left[\Pp\, \dot{\X} + e( \mathbb{P}^2 + m^2 \mathbb{I} )\right],
   \end{align}
   where we have also used~\eqref{eq:Pmat} and~\eqref{eq:Pmat1}. Since for $N=1$, $\mathbb{P}^2=p^ap_a \mathbb{I}$ we can even write the very same action in the form 
   \begin{align}
   \label{eq:canucmat}
       S_{\text{Ham}}^{\text{unc.}} [\X,\Pp,\mathbb{L}]= \frac12 \int d\tau \Tr \left[\Pp\, \dot{\X} + \mathbb{L}( \mathbb{P}^2 + m^2 \mathbb{I} )\right],
   \end{align}
   where $\mathbb{L}$ is now a matrix of Lagrange multipliers as introduced in~\eqref{eq:Lmat} with $\frac12\Tr\, \mathbb{L}=e$. 
   \item
   \underline{Orbit action}: 
   \begin{align}
   \label{eq:Lorbuc}
     S_{\text{orb}}^{\text{unc.}} [x^a, \vec{p}\,]= \int d\tau \,\dot{x}^a\, p_a\,,\quad\quad  p_0 = \pm \sqrt{\vec{p}^{\,2}+m^2}\,,
   \end{align}
   where the condition $p_0 = \pm \sqrt{\vec{p}^{\,2}+m^2}$ selects only one massive orbit, depending on the sign one chooses. This action is in a sense the simplest action one can write for any given co-adjoint orbit, here again restricted to semi-simple orbits since $p^ap_a+m^2=0$ by construction. This action also follows from the method of non-linear realisation as we discuss in more detail in appendix~\ref{app:FP}.
\end{enumerate}

The number of degrees of freedom can be determined from the actions in all three cases and one finds two degrees of freedom (in configuration space). This number of degrees of freedom agrees also with the dimension of the orbit of a massive momentum under $SO(1,2)$. Even though there are two distinct orbits, corresponding to hyperboloids in the forward and backward light-cone or positive and negative energy particles, respectively, the dimension of all these orbits is the same.\footnote{All these actions take exactly the same form in $D$-dimensional space-time, where the massive particle has $D-1$ degrees of freedom.}

For the case of the massless particle one can most easily proceed from the Hamiltonian action~\eqref{eq:LHamuc} which is perfectly well-behaved in the limit $m\to 0$. The mass-shell constraint $p^a p_a=0$ now selects nilpotent orbits in momentum space, corresponding to the light-cones themselves. Imposing the corresponding orbit condition $p_0=\pm \sqrt{\vec{p}^{\,2}}$ in~\eqref{eq:Lorbuc} describes the same physics.  

Thinking of an (irreducible) particle as corresponding to a single momentum orbit, the most natural action is~\eqref{eq:Lorbuc} as it only contains a single orbit whereas both~\eqref{eq:Lgeouc} and~\eqref{eq:LHamuc} collect several orbits together.
While for the uncoloured Poincar\'e algebra, there are only two distinct massive orbits (of positive and negative energy) and the distinction may seem innocuous, we shall see next that for the coloured Poincar\'e algebra the distinction is more important.

\subsubsection{Coloured Minkowski space}

\begin{enumerate}
    \item 
    \underline{Geometric action}:
We have determined the invariant line element in~\eqref{eq:invmetric} and  can therefore immediately write a world-line reparametrisation invariant action for a massive particle 
\begin{align}
\label{eq:geoL}
    S_{\text{geo}}[\X]] = - m \int  d\tau 
    \sqrt{ - \tfrac1{2N}\Tr\big[ \dot{\X}^2 \big]} \,.
\end{align}
The conjugate momentum is given by
\begin{align}
\Pp=  \frac{1}{2N} m\frac{\dot{\X}^T}
{\sqrt{-\tfrac1{2N}\Tr\big[{\dot\X}^2}]}
\end{align}
that satisfies the constraint 
\begin{align}
\label{trconstraint}
\Tr\big[{\Pp}^2+m^2
\mathbb{I}]=0\,,
\end{align}
where $\mathbb{I}$ is the $2N\times 2N$ unit matrix.
This scalar constraint is the generator of world-line diffeomporphisms. The time-like nature of the coloured velocity corresponds to $-\Tr\big[\dot{\X}^2\big]>0$. The equations of motion implied by this geometric action in proper time gauge are simply
\begin{align}
\label{eq:geoEOM}
    \ddot{\X} = 0\,.
\end{align}
The proper time gauge condition implies $-\tfrac{1}{2N}\Tr\big[\dot{\X}^2\big]=1$.
\item
    \underline{Hamiltonian action}:
    In order to write the Hamiltonian action we now have several expressions that could be generalised. The canonical action that corresponds directly to the geometric action above is
\begin{align}
    \tilde{S}_{\text{Ham}} [\X,\Pp,e] =\frac{1}{2N}\int d\tau\, \Tr\big[\Pp\,\dot{\X}+ e \,(\Pp^2+m^2\,\mathbb{I})\big]\,,
    \label{eq:GeometricActionH}
\end{align}
and it is the direct generalisation of~\eqref{eq:canuc}. Integrating out $\Pp$ and $e$ from this action is not straight-forward and we discuss some aspects of this in appendix~\ref{app:HamLag}.
By varying~\eqref{eq:GeometricActionH} with respect to $e$, we re-obtain the scalar constraint~\eqref{trconstraint}. There are many solutions to this scalar constraint, corresponding to 
a continuum of different semi-simple orbits. These orbits can even have different dimensions and thus describe particles with different numbers of degrees of freedom. We discuss aspects of these orbits in more detail below in section~\ref{sec:orbits}.

In order to always have massive particles with the same number of degrees of freedom, we can alternatively consider the generalisation of~\eqref{eq:canucmat} that becomes
\begin{align}
\label{eq:canHmat}
  S_{\text{Ham}}[ \X,\Pp,\mathbb{L}]=\frac{1}{2N}\int d\tau\, \Tr\big[\Pp\,\dot{\X}+ \mathbb{L} \,(\Pp^2+m^2\,\mathbb{I})\big]\,,
\end{align}
now with a matrix-valued Lagrangian multiplier $\mathbb{L}$ as written out in~\eqref{eq:Lmat}. Note that $\ci\,\mathbb{L}\in \mf{u}(N,N)$ and therefore there is a non-vanishing component $\tfrac{1}{2N}\Tr\big[\mathbb{L}\big]=e$. The constraint that the matrix-valued Lagrange multipliers imposes is
\begin{align}
\label{eq:matconst}
    \Pp^2+m^2\,\mathbb{I}  = 0
\end{align}
which is a $(2N)\times(2N)$ matrix condition. The scalar constraint~\eqref{trconstraint} is its trace and therefore a weaker condition; only in the case $N=1$ the conditions are equivalent by virtue of~\eqref{eq:Pmat1}. \\
The matrix constraint~\eqref{eq:matconst} restricts the possible momentum orbit more strongly than the trace constraint and, in particular, all semi-simple orbits satisfying~\eqref{eq:matconst} have the same dimension as we shall review in more detail in section~\ref{sec:orbits}.
\item
  \underline{Orbit action}:
    In analogy with~\eqref{eq:Lorbuc}, we can also consider the simple action associated with a given orbit. This corresponds to considering
    \begin{align}
    \label{eq:orbact}
        S_{\text{orb}} = \frac{1}{2N} \int d\tau \Tr\big[  \Pp\,\dot{\X}\big] = \int d\tau \langle \Pp, \dot{\X}\rangle \,,\quad\quad
        \text{$\Pp$ belonging to a given massive orbit.}
    \end{align}
In the second step we have introduced the canonical pairing between $\mf{su}(N,N)$ and its dual:
for any element $\cB\in \mf{su}(N,N)^*$\,, there is a unique element
$H_{\cB}\in \mf{su}(N,N)$ satisfying
\be
	\la \cB, \,A \ra=\frac1{2N}\,\Tr\left[A\,H_{\cB}\right],
	\qquad \text{for all }  A\in \mf{su}(N,N)\,,
	\label{dual}
\ee
where the trace is for the fundamental representation of $\mf{su}(N,N)$. By abuse of notation, we write $\mathbb{P}$ both for the adjoint and the coadjoint representative.

We briefly explain in this case why the orbit action agrees with the action one would derive from a standard non-linear realisation~\cite{Coleman1,Coleman2,Ogievetski,Gomis:2006xw} for a semi-direct product. We do this for simplicity by focussing on the case when a representative of the massive orbit is given by
\begin{align}
\label{eq:rep1}
    \Pp= m \mathcal{P}_0^0 = m \begin{pmatrix} \ci \mathbf{I} &0\\0& -\ci \mathbf{I}\end{pmatrix},
\end{align}
where $\mathcal{P}_0^0\in \mf{su}(N,N)^*$ is the generator dual to $P_0^0$, the usual momentum. The stabiliser of this momentum is generated by 
\begin{align}
\label{eq:stabm}
M_0\equiv M_0^0 \,,\quad M_0^I\,,\quad N^I
\end{align}
as can be checked using the algebra~\eqref{eq:CPoin}. The algebra generated by~\eqref{eq:stabm} is $\mf{u}(1)\oplus\mf{su}(N)\oplus \mf{su}(N)$. 
We thus take $H=\left(U(1)\times SU(N)\times SU(N) \right)$ 
and write a local representative of the coset in the form
\begin{align}
\label{eq:cosel}
g = e^{\X}\,b\, h \,,
\end{align}
with 
\begin{subequations}
\begin{align}
\X &= x^a P_a + x^a_I P_a^I + y_I Q^I\quad &&\textrm{(generalised coordinate of coloured Minkowski space)}\,,\\
\label{eq:genb}
b &= \exp\left[ v^i M_i + w^i_I M_i^I \right] \quad &&\textrm{(generalised boost)}\,,
\end{align}
\end{subequations}
and $h$ belonging to the generalised little group $H$ generated by~\eqref{eq:stabm}. The indices $a=0,1,2$ are covariant while $i=1,2$ are purely spatial indices. The generalised boost $b$ represents the broken generalised Lorentz generators acting on the generalised coordinate $\X$. We think of $x^a$ as the uncoloured position variable and $x^a_I$ and $y_I$ as coloured generalisation required by the algebra in the way described in section~\ref{sec:colMink}. All coordinates here are real.

The Maurer--Cartan form associated with the coset representative~\eqref{eq:cosel} is
\begin{align}
\Omega = g^{-1} dg = b^{-1} d\X b+ b^{-1} db
\end{align}
where we have chosen the gauge $h=1$. The component of $\Omega$ along $P_0$ is invariant under the little group
and its pullback to the world-line can be used as a Lagrangian that by construction then will be invariant under the full coloured Poincar\'e group. Using the pairing with the dual space of the coloured translation
\begin{align}
L_{\text{nlr}}\, dt  = m  \Big[\Omega^*\Big]_{P_0} = \langle m\,\mathcal{P}_0,b^{-1}\, 
d\X\, b\rangle\,,
\end{align}
where $\langle \cdot, \cdot \rangle$ is the pairing of the Lie algebra with its dual and $\mathcal{P}_0$ is the dual of the generator $P_0$. Now, using the invariance of the pairing, we can rewrite this as
\begin{align}
L_{\text{nlr}}  = \langle  \Ad^*_b (m\,\mathcal{P}_0),\tfrac{d\X}{dt} \rangle = \langle  \Pp,\dot\X\rangle\,,
\end{align}
where now $\Pp$ is any element of the orbit of the reference momentum.\footnote{Note that
$\mathbb P=\Ad^*_b (m\,\mathcal{P}_0)$
belongs to a `momentum' orbit,
which is 
different from the coadjoint orbit, $\{\Ad^*_g (m\,\mathcal{P}_0)\,|\,\forall g\in 
\CPoin_3(N)\}$,  of coloured
Poincar\'e symmetry $\CPoin_3(N)$.
The coadjoint orbit can be viewed 
as the `phase space',
 and hence the momentum orbit
 is its Lagrangian subspace.
Because the (coloured) Poincar\'e symmetry is not semi-simple,
the coadjoint orbits 
are not isomorphic to
the adjoint orbits.
In fact the momentum orbit
can be viewed as either 
the adjoint orbit of $\CPoin_3(N)$
or the coadjoint orbit
of the coloured Lorentz
group $SU(N,N)$.
}
We thus have shown that the free particle Lagrangian obtained from the non-linear realisation is equal to that of the orbit construction~\eqref{eq:orbact}, see also appendix~\ref{app:FP}.

\end{enumerate}

\subsection{\texorpdfstring{Coadjoint orbits of $\mf{su}(N,N)$}{Coadjoint orbits of su(N,N)}}
\label{sec:orbits}

The coloured Lorentz group $SU(N,N)$ acts on the momentum $\Pp\in \mf{su}(N,N)^*$ using the coadjoint action. In the above actions we have seen that the matrix constraint
\begin{align}
    \Pp^2 + m^2\mathbb{I} =0
\end{align}
and the scalar constraint 
\begin{align}
\Tr \big[\Pp^2 + m^2\mathbb{I}\big] =0
\end{align}
arose. These constraints are both $SU(N,N)$ covariant and the solutions can therefore be classified in terms of coadjoint orbits. Since $\mf{su}(N,N)$ is simple we can identity coadjoint orbits with adjoint orbits using the non-degenerate trace pairing~\eqref{dual}. For a general discussion of orbits see~\cite{CollingwoodMcGovern}.

\subsubsection{Massive particles: semi-simple orbits}

For $m>0$, the relevant orbits are of  semi-simple type and have representatives of the form
\begin{align}
    \Pp = \text{diag} ( a_1,\ldots, a_N, b_1\ldots, b_N)\,.
\end{align}
The condition $\Pp \in \mf{su}(N,N)^*$ implies that $\sum_{j=1}^n (a_j + b_j)=0$ and our hermiticity conditions~\eqref{2N rep} imply that the $a_j$ and $b_j$ are pure imaginary. Interchanging $a_j\leftrightarrow a_k$ or $b_j\leftrightarrow b_k$ for the representative does not change the orbit.

Imposing the matrix constraint implies that $a_j=b_j=\pm \ci m$ for all $j=1,\ldots,N$ while the scalar constraint implies the much weaker condition $\sum_{j=1}^N (a_j^2 + b_j^2) = -2N m^2$.

As a representative solution of the matrix constraint we have already given~\eqref{eq:rep1}. Its stabiliser was determined to have the Lie algebra $\mf{u}(1)\oplus \mf{su}(N)\oplus \mf{su}(N)$ and thus the size of the corresponding orbit $\mathcal{O}_{\eqref{eq:rep1}}$ is
\begin{align}
    \dim \mathcal{O}_{\eqref{eq:rep1}} &= \dim \mf{su}(N,N) - \dim (\mf{u}(1)\oplus \mf{su}(N)\oplus \mf{su}(N)) = (2N)^2-1 - (1+ 2 (N^2-1)) \nn\\
    & = 2N^2
\end{align}
by applying the orbit-stabiliser theorem. The same is true for any other orbit satisfying the matrix constraint~\eqref{eq:matconst}.

By contrast, if only the scalar constraint~\eqref{trconstraint} is satisfied, there are many more solutions. Consider for example
\begin{align}
\label{eq:rep2}
    \Pp = \sqrt{N}\, \text{diag}(\ci m,0,0,\ldots,0,-\ci m)\,,
\end{align}
which solves the trace but not the matrix constraint.
The stabiliser is in this case generated by $\mf{u}(1)\oplus \mf{u}(N-1,N-1)$,
so that the orbit is of dimension
\begin{align}
    \dim \mathcal{O}_{\eqref{eq:rep2}} &= \dim \mf{su}(N,N) - \dim (\mf{u}(1)\oplus \mf{u}(N-1,N-1)) = 8N-6
\end{align}
and is a smaller orbit than the one dictated by the matrix constraint for $N>3$.

As a final example we consider
\begin{align}
\label{eq:rep3}
    \Pp = \sqrt{\frac{6}{(2N+1)(N+1)}} \,\text{diag}(\ci m,2 \ci m,\ldots, N \ci m, -N\ci m,\ldots,-2\ci m,-\ci m)\,,
\end{align}
which solves the trace but not the matrix constraint. Its stabiliser is $\mf{u}(1)^{2N-1}$ and the orbit has dimension
\begin{align}
    \dim \mathcal{O}_{\eqref{eq:rep3}} &= \dim \mf{su}(N,N) - \dim (\mf{u}(1)^{2N-1}) = 2N(2N-1)\,.
\end{align}
This is the maximal dimension of a semi-simple orbit since the stabiliser is the minimal one: just the Cartan subalgebra. The dimension of the orbit equals the number of roots of $\mf{su}(N,N)$ and is much larger than the orbits corresponding to the matrix constraint for $N>1$.

Generally, (complex) semi-simple orbits are related to representative elements in the closure of the fundamental Weyl chamber (within the Cartan subalgebra)~\cite[\S2]{CollingwoodMcGovern}. Fully generic elements in the interior of the fundamental Weyl chamber correspond to elements such as the example~\eqref{eq:rep3} and the corresponding semi-simple orbits are of maximal dimension. If the semi-simple element is non-generic and has a larger stabiliser, it lies on boundaries of the fundamental Weyl chamber and the dimension of the orbit shrinks. The matrix constraint~\eqref{eq:matconst} selects a very specific set of such boundaries where all orbits have the same dimension $(2N)^2$. As is clear from the examples above, the trace constraint~\eqref{trconstraint} allows for a wide variety of orbits. In fact, from a projective point of view all possible semi-simple orbits are allowed. 

Thinking of different semi-simple orbits as different types of massive particles, just like positive and negative energy particles in the uncoloured case, we conclude that the geometric action~\eqref{eq:geoL} and the canonical action~\eqref{eq:GeometricActionH} contain all possible types of massive particles, where even the number of degrees of freedom can change.

By contrast, the canonical action~\eqref{eq:canHmat} only contains massive particles with $(2N)^2$ degrees of freedom and contains all orbits of this dimension.

The orbit action~\eqref{eq:orbact} describes only a single orbit and the choice of a type of massive particle corresponds to a choice of semi-simple orbit.

\medskip

For completeness, we give the general form of all representatives of solutions of the matrix constraint~\eqref{eq:matconst}.
By using $SU(N,N)$ transformations, the diagonal solution for $\mathbb P^2+m^2\,\mathbb I=0$ can be brought into the form
\be
	\widehat{\mathbb P}_{N-2\ell}=m\,{\rm diag}(\underbrace{\ci,\ldots, \ci}_{\ell}, \underbrace{-\ci, \ldots, -\ci}_{N-\ell},
	\underbrace{\ci, \ldots \ci}_{N-\ell},\underbrace{-\ci, \ldots, -\ci}_\ell)\,,
	\label{repres}
\ee
where we have taken into account the equivalence of interchanging diagonal elements among the first and the second $N$ entries respectively, since such interchanges can be made by
an $SU(N,N)$ action.
Here, the number $\ell$ is the number of the element $+\ci$ in the first $N$ entries. Hence, $\ell$ can take
$N+1$ possible values ($\ell=0,1,\ldots, N$) and 
there exist $N+1$ distinct orbits satisfying the equation $\mathbb  P^2+m^2\,\mathbb I=0$\,.
The dimension of the orbit ${\cal O}_{N-2\ell}$ generated by  
$\widehat{\mathbb P}_{N-2\ell}$  is simply $2N^2$,
and it corresponds to the
homogeneous space,
\be 
\mathcal{O}_{N-2\ell}
\simeq U(N,N)/(U(N-\ell,
\ell)\times U(N-\ell,\ell))\,,
\ee 
which is a pseudo-Euclidean flag variety.
The stabiliser is compact only for $\ell=0$ and $\ell=N$.
Note that the dimension does not depend on $\ell$.
As mentioned previously,
the orbits that we describe here correspond to semi-simple coadjoint orbits of $\mathfrak{su}(N,N)$
(and they are related by Weyl reflections).

\subsubsection{Massless particles: nilpotent orbits}

For the case $m=0$, both the matrix and the trace constraint require the orbit to have a nilpotent representative. Nilpotent orbits of $SU(N,N)$ are described by signed Young tableaux~\cite[Thm.~9.3.3]{CollingwoodMcGovern}.  These are 
Young diagrams with $2N$ boxes filled with $N$ positive and $N$ negative signs
in such a way that signs alternate across rows.
Two signed Young diagrams are regarded as equivalent if and only if 
one can be obtained from the other by interchanging rows of equal length.
Say, the height of the $k$-th column is $h_k$, then 
the dimension of the orbit corresponding to the Young diagram ${\bf \lambda}=[h_1,\ldots, h_{2N}]$ is
\begin{align}
	\dim\, {\cal O}_{\bf\lambda}=4\,N^2-\sum_{k=1}^{2N}\,h_k^2\,. 
\end{align}
The orbits with dimension $2 N^2$ correspond to the Young diagram ${\bf\lambda}=[N,N]$
and there are $N+1$ different signed Young diagrams:
\begin{align}
	\young(+-,::,+-,-+,::,-+)\,.
	\label{young}
\end{align}
These are the nilpotent orbits which are the $m\to 0$ limits of the semi-simple orbits described by the matrix constraint $\mathbb P^2=-m^2\,\mathbb I$\,,
and they obviously satisfy the orbit equation $\mathbb P^2=0$\,.
However, the algebraic variety given by the latter equation does not contain only the $2N^2$ dimensional nilpotent orbits,
but also other nilpotent orbits of smaller dimensions. 
Such orbits correspond to all signed Young diagrams of the type ${\bm \lambda}=[2N-k,k]$ 
(which can be obtained from the Young diagram \eqref{young} by moving boxes from the second to the first column).
The number of such signed Young diagrams is $k+1$
and its dimension is $4N^2-(2N-k)^2-k^2=4Nk-2k^2$\,.
The number $k$ takes value in $\{0,\ldots, N\}$
including the $2N^2$ dimensional one with $k=N$,
so the equation $\mathbb P^2=0$ describes 
$\sum_{k=0}^{N}(k+1)=(N+1)(N+2)/2$ distinct orbits including the trivial one $\mathbb P=0$. The massless particles associated to these orbits are discussed in section~\ref{sec:massless}.

\subsection{Massive coloured particle in component form}

We now work out the Lagrangians in more detail, beginning with the more general one~\eqref{eq:canHmat} that contains all massive orbits satisfying the matrix-valued `mass-shell' constraint~\eqref{eq:matconst}.

Plugging the parametrisations~\eqref{eq:Xmat},~\eqref{eq:Pmat} and~\eqref{eq:Lmat} into the action~\eqref{eq:canHmat} leads to the Lagrangian
\begin{align}
\label{eq:L2n}
L = \frac{1}{N} \tr \bigg[
\dot{\bf X}^a\, {\bf P}_a -\dot{\bf Y}\,{\bf \Pi} + {\bf \Lambda} \Big( {\bf P}_a {\bf P}^a -{\bf \Pi}^2 +m^2 \id\Big)
+ {\bf K}_a \Big( \{ {\bf P}^a, {\bf \Pi} \} + \frac{\ci}{2} \varepsilon^{abc} [ {\bf P}_b, {\bf P}_c ] \Big)
\bigg]\,,
\end{align}
where our space-time conventions were given in footnote~\ref{fn:eps}.

Using the parametrisations~\eqref{eq:Xpar},~\eqref{eq:Ppar} and~\eqref{eq:Lpar}, the Lagrangian~\eqref{eq:L2n}  becomes 
\begin{align}
\label{eq:L2comp}
L & = \dot{x}^a \,p_a +\frac1{N} \,\dot{x}^a_I \,p_{a}^I - \frac1{N} \,\dot{y}^I\, \pi_I +  e\Big(p^a\,p_a +\frac{1}{N}\,p^a_I\, p_{aI} -\frac{1}{N}\,\pi^I \,\pi_I + m^2 \Big)\nn\\
&\quad  + \frac{1}{N}\,\lambda_I \Big( 2\, p^a\, p_a^I -\frac{1}{2} \,d^{JKI} p^a_{J} \,p_{aK}  - \frac{1}{2} d^{JKI}\, \pi_J\, \pi_K \Big)\nn\\
&\quad + \frac{2}{N} \,f_a\, p^a_I\, \pi^I + \frac{1}{N}\,k_{aI} \Big( 2 \,p^a \,\pi^I -  d^{JKI}\, p_{aJ} \,\pi_K + \frac12 \,\varepsilon^{abc}\, f^{JKI}\, p_{bJ} \,p_{cK}\Big)\,,
\end{align}
when we also use the algebra~\eqref{eq:unass}. The adjoint  $\mf{su}(N)$ indices $I$ are raised and lowered with the invariant $\delta_{IJ}$. 

Focussing on the terms that do not depend on $N$, we recognise immediately the usual uncoloured~\eqref{eq:LHamuc}. The terms proportional to $1/N$ represent the effects of colouring the Poincar\'e algebra.

We record the equations of motion implied by~\eqref{eq:L2comp}
\begin{subequations}
\label{eq:eoms}
\begin{align}
    \dot{p}_a =0 \,, \quad\quad  \dot{p}_a^I =0 \,,\quad\quad \dot{\pi}_I &=0\,,\\
    \dot{x}^a + 2 e p^a +\frac{2}{N} \lambda_I p^{a I} +\frac{2}{N} k_{a I}\pi^I &=0\,,\\
   \dot{x}^a_I + 2e p^a_I +2 \lambda_I p^a - d^{IJK} \lambda_J p_{a K} + 2f_a\pi^I - d^{IJK} k_{a J}\pi_K + \varepsilon_{abc} f^{IJK} k^b_J p^c_K &=0\,,\\
   -\dot{y}^I - 2e \pi^I - d^{IJK} \pi_J\pi_K +2 f_a p^{aI} +2 k_a^I p^a -d^{IJK} k_{aJ} p^a_K &=0\,,\\
   \label{eq:eeq}
  p^a p_a +m^2 +\frac{1}{N} p^a_I p_a^I-\frac1{N} \pi^I \pi_I &=0\,,\\
  \label{eq:lIeq}
2p_a^{I} p^a -\frac12 d^{IJK} p_{aJ} p^a_K - \frac12 d^{IJK} \pi_J\pi_K &=0 \,,\\
    \label{eq:faeq}
p_a^I \pi_I &= 0\,,\\
\label{eq:kaIeq}
  2p^a \pi^I - d^{IJK} p_{J}^a \pi_K +\frac12 \varepsilon^{abc} f^{IJK} p_{bJ} p_{cK}&=0\,.
\end{align}
\end{subequations}

The constraints that are enforced by the Lagrange multipliers are given in~\eqref{eq:eeq}--\eqref{eq:kaIeq}. The total number of constraints is $1+(N^2{-}1)+3+3(N^2{-}1)=(2N)^2$ since the Lagrange multiplier $\mathbb{L}$ belongs to $\mf{u}(N,N)$.
We note that these constraints are not all independent and we are dealing with a reducible system of constraints. This will be further investigated in section~\ref{sec:redconst}.

\medskip

While the Lagrangian~\eqref{eq:L2comp} is valid for any choice of $\mathbb{P}$ satisfying $\mathbb{P}^2=-m^2\,\mathbb{I}$, the orbit Lagrangian~\eqref{eq:orbact} depends on the choice of a given orbit where this condition is satisfied. In order to investigate this we take a representative momentum and repeat the analysis from appendix~\ref{app:FP} in connection with the non-linear realisation discussed above.

Written in matrix form, the representative momentum~\eqref{eq:rep1} takes the form 
\begin{align}
\label{eq:MPref}
\widehat{\mathbb{P}} = m \begin{pmatrix} \ci {\bf I}& 0 \\ 0 & -\ci {\bf I}\end{pmatrix}
\end{align}
while the boost element $b$ from~\eqref{eq:genb} can be written as
\begin{align}
b = \exp\left[\frac12  \begin{pmatrix}
0 & {\bf V}^+ \\ {\bf V}^- &0
\end{pmatrix}\right]
\,,\quad\text{with} \quad({\bf V}^+)^\dagger = {\bf V}^-\,.
\end{align}
Other elements of $\mf{su}(N,N)$ stabilise $\widehat{\mathbb{P}}$ under the adjoint action and belong to the subalgebra $\mf{u}(1)\oplus \mf{su}(N)\oplus \mf{su}(N)$.

The orbit of the reference momentum is then given by
\begin{align}
b \,\widehat{\mathbb{P}} \,b^{-1} = m \begin{pmatrix}
\ci \cosh \sqrt{ {\bf V}^+ {\bf V}^-} & -\ci \frac{\sinh \sqrt{{\bf V}^+{\bf V}^-}}{\sqrt{{\bf V}^+{\bf V}^-}} {\bf V}^+\\
\ci {\bf V}^- \frac{\sinh \sqrt{{\bf V}^+{\bf V}^-}}{\sqrt{{\bf V}^+{\bf V}^-}}  & -\ci \cosh \sqrt{ {\bf V}^- {\bf V}^+}
\end{pmatrix}\,,
\end{align}
where the hyperbolic trigonometric functions of the hermitian matrices ${\bf V}^+{\bf V}^-$ and ${\bf V}^-{\bf V}^+$ are defined by their power series expansions. The formula represents the action of a general coloured Lorentz boost on the reference momentum~\eqref{eq:MPref}.
We rewrite this more compactly by defining
\begin{align}
{\bf P}^+  =  - m\, \ci \frac{\sinh \sqrt{{\bf V}^+{\bf V}^-}}{\sqrt{{\bf V}^+{\bf V}^-}} {\bf V}^+\,,\quad\quad
{\bf P}^- = \left({\bf P}^+\right)^\dagger\,,
\end{align}
leading to
\begin{align}
\label{eq:MPorb}
b \,\widehat{\mathbb{P}} \,b^{-1} =   \begin{pmatrix}
\ci \sqrt{ m^2\, {\bf I}+ {\bf P}^+ {\bf P}^- }  & {\bf P}^+\\
{\bf P}^- & -\ci \sqrt{m^2\, {\bf I} + {\bf P}^- {\bf P}^+ }
\end{pmatrix}\,,
\end{align}
where the square root of matrices is again defined by its Taylor series around ${\bf P}^{\pm}{\bf P}^{\mp}=0$. We note that all elements in the orbit of the reference momentum~\eqref{eq:MPref} satisfy
\begin{align}
 \left( b \,\widehat{\mathbb{P}} \,b^{-1} \right)^2=   \widehat{\mathbb{P}}^2 = - m^2 \mathbb{I}\,.
\end{align}
This equation has more solutions than~\eqref{eq:MPorb} and the Lagrangian~\eqref{eq:canHmat} captures all of them.

Considering the Lagrangian~\eqref{eq:orbact} for the reference momentum~\eqref{eq:MPref} we need to pair ~\eqref{eq:MPorb} with the velocity vector $\dot{\mathbb{X}}$ given in~\eqref{eq:Xmat}, which leads to
\begin{align}
\label{eq:L1cexp1}
L_{\text{orb}} = \frac{1}{2N} \tr \bigg[  ( \dot{{\bf X}}_0  -  \dot{{\bf Y}})  \sqrt{m^2\, {\bf I} + {\bf P}^+{\bf P}^-}  +( \dot{{\bf X}}_0  +  \dot{{\bf Y}}) \sqrt{m^2\, {\bf I} + {\bf P}^-{\bf P}^+}  + \dot{{\bf X}}^+ {\bf P}^- + \dot{{\bf X}}^- {\bf P}^+\bigg]\,.
\end{align}
This is the coloured generalisation of the Lagrangian~\eqref{eq:Lorbuc}. Setting $N=1$ and $\dot{{\bf X}}^a$, ${\bf P}^a$ to 3-component Lorentz vectors  and $\dot{{\bf Y}}=0$ reduces it manifestly to the standard uncoloured massive  particle.

The Lagrangian~\eqref{eq:L1cexp1} has been deduced from the representative momentum~\eqref{eq:MPref}. We shall now explore the relation to the previously derived Lagrangian~\eqref{eq:L2n} and try to follow a route similar to that of appendix~\ref{app:FP} that starts from the constraints implied by~\eqref{eq:L1cexp1}. 
Varying~\eqref{eq:L1cexp1} with respect to ${\bf X}^0\pm {\bf Y}$ leads to the constraints
\begin{align}
\label{eq:C1}
{\bf C}_\pm = {\bf P}^0 \pm {\bf \Pi} - \sqrt{m^2 {\bf I} + {\bf P}^\pm {\bf P}^{\mp}} = 0 
\end{align}
that are $(N\times N)$-matrix valued and where we have introduced new $(N\times N)$-matrix valued variables ${\bf P}^0=-{\bf P}_0$ and ${\bf \Pi}$ as integration constants.  These constraints are the generalisations of the positive-energy mass-shell constraint of an uncoloured particle as in~\eqref{eq:Lorbuc}. In order to allow for generalisations of negative energy, we shall consider as a generalisation the constraints
\begin{align}
\label{eq:C2}
{\bf C}_\pm^2 = \left({\bf P}^0\pm {\bf \Pi}\right)^2-\left(m^2 {\bf I}+ {\bf P}^{\pm} {\bf P}^{\mp} \right)=0 \,.
\end{align}
These are also $(N\times N)$-matrix valued and will be enforced using the matrix-valued Lagrange multipliers ${\bf \Lambda}\pm {\bf K}$. Enforcing only these squared constraints allows for more solutions than implied by the original Lagrangian in the same way as squaring the positive mass condition for going from~\eqref{eq:Lorbuc} to~\eqref{eq:LHamuc} allows for more solutions, see also appendix~\ref{app:FP}.

The resulting generalised Lagrangian is then
\begin{align}
\label{eq:L2}
\tilde{L} &= \frac1{2N} \tr\bigg[-  (\dot{\bf X}^0 + \dot{\bf Y}) \left({\bf P}^0+ {\bf \Pi}\right)- (\dot{\bf X}^0 - \dot{\bf Y}) \left({\bf P}^0- {\bf \Pi}\right) + {\bf P}^+ \dot{\bf X}^- + {\bf P}^- \dot{\bf X}^+\nn\\
&\quad - \left({\bf \Lambda}+ {\bf K}\right)\left(\left({\bf P}^0 + {\bf \Pi}\right)^2-\left(m^2{\bf I}+ {\bf P}^+ {\bf P}^- \right)\right)
- \left({\bf \Lambda}- {\bf K}\right)\left(\left({\bf P}^0- {\bf \Pi}\right)^2-\left(m^2{\bf I} + {\bf P}^-{\bf P}^+ \right)\right)
\bigg]\nn\\
&= \frac1{N} \tr\bigg[\dot{\bf X}^a {\bf P}_a -  \dot{\bf Y} {\bf \Pi} +  {\bf \Lambda} \left(   {\bf P}^a{\bf P}_a - {\bf \Pi}^2 +m^2 {\bf I}\right) 
- {\bf K} \left(   \left\{ {\bf P}^0, {\bf \Pi} \right\} +  \ci \left[ {\bf P}^1,{\bf P}^2\right]
\right)
\bigg]
\end{align}
with ${\bf P}^\pm = {\bf P}_1 \pm \ci {\bf P}_2= {\bf P}^1 \pm \ci {\bf P}^2$ as well as ${\bf P}_0=-{\bf P}^0$ due to the $(-++)$ signature.

The Lagrangian~\eqref{eq:L2} does not manifestly agree with~\eqref{eq:L2n} that we deduced from the natural squared constraint~\eqref{eq:matconst} in $\mf{u}(N,N)$, in particular, it does not exhibit manifest Lorentz invariance under the usual $SO(1,2)$ group. 
However, the transformation of the Lagrange multipliers under the coloured Lorentz group is exactly such that it is opposite to that of the constraints so that the Lagrangian is invariant by construction. 

The matrix mass-shell Lagrangian~\eqref{eq:L2n} also has two extra Lagrange multipliers ${\bf K}_1$ and ${\bf K}_2$ that are not present in~\eqref{eq:L2}. The constraints they enforce from~\eqref{eq:L2n} are 
\begin{align}
 \left\{ {\bf P}^1, {\bf \Pi} \right\} +  \ci \left[ {\bf P}_2,{\bf P}_0\right] = 0 \,,\quad
  \left\{ {\bf P}^2, {\bf \Pi} \right\} + \ci  \left[ {\bf P}_0,{\bf P}_1\right] = 0
\end{align}
that are equivalent to 
\begin{align}
\left({\bf P}_0 \pm {\bf \Pi}\right)  {\bf P}^\pm  = {\bf P}^{\pm}  \left( {\bf P}_0 \mp {\bf \Pi}\right)\,.
\end{align}
When the original constraint~\eqref{eq:C1} is satisfied, meaning ${\bf P}_0 \pm {\bf \Pi} = \sqrt{m^2 {\bf I} + {\bf P}^\mp {\bf P}^\pm}$, the above constraints follow. However, they need not follow from the squared constraint ${\bf C}_\pm^2=0$ in~\eqref{eq:C2} and thus the manifestly covariant Lagrangian~\eqref{eq:L2n} is not equivalent to~\eqref{eq:L2}. We therefore consider the manifestly covariant Lagrangian~\eqref{eq:L2n} as the appropriate generalisation of the massive particle~\eqref{eq:MPref} to all possible massive orbits and shall work with~\eqref{eq:L2n} in the following.

\subsection{Constraint structure and degrees of freedom}
\label{sec:redconst}

In this section we revisit the discussion of the dimension of semi-simple orbits of section~\ref{sec:orbits} and relate it to the reducibility of the constraints implied by the matrix constraint~\eqref{eq:matconst}.
The reducibility of the constraint means that not all of the $(2N)^2$ components of~\eqref{eq:matconst} can be counted independently. Rather, we are interested in the dimension of the solution space of $\Pp^2=-m^2 \mathbb{I}$. 

{}From the discussion in section~\ref{sec:orbits} we know that the dimension of the solution space is given by the size of an orbit of solutions and that these can be characterised by the stabilisers of the semi-simple representative of the orbit. For the case of the massive momentum given in~\eqref{eq:rep1} the stabiliser was $S(U(N)\times U(N)) \subset SU(N,N)$. For any other orbit of solutions of~\eqref{eq:matconst} the stabiliser is a similar real form of the same dimension. 
The dimension of the stabiliser equals the number of independent constraints contained in~\eqref{eq:matconst} since this is the number by which the number of variables is reduced to obtain an orbit of solutions. 

A canonical analysis then reveals the number of degrees of freedom. It is important to consider the configuration space variables of~\eqref{eq:canHmat} as $\X$, $\mathbb{P}$ and $\mathbb{L}$, where the first two have $(2N)^2-1$ components while $\mathbb{L}$ has $(2N)^2$ components. Canonically, the conjugate momenta of $\mathbb{P}$ and $\mathbb{L}$ vanish and there is another constraint relating the conjugate momentum of $\X$ to $\mathbb{P}$. These are all the primary constraints and their numbers are $(2N)^2-1$, $(2N)^2$ and $(2N)^2-1$, respectively.

The constraint~\eqref{eq:matconst} appears as a secondary constraint after time evolution of the primary constraint that the momentum $\pi_{\mathbb{L}}$ conjugate to $\mathbb{L}$ vanishes. As we argued above, there are only $2N^2-1$ independent constraints contained in~\eqref{eq:matconst} in agreement with the dimension of the stabiliser of a solution. These $2N^2-1$ constraints and the $(2N)^2$ conditions $\pi_{\mathbb{L}}=0$ are first-class constraints, the remaining two sets of constraints are second-class. The count of degrees of freedom in phase space is therefore
\begin{align}
    \underbrace{2{\times}( (2N)^2-1)}_{(\X,\pi_{\X})} + \underbrace{2{\times}( (2N)^2-1)}_{(\mathbb{P},\pi_{\Pp})} +\underbrace{2{\times} (2N)^2}_{(\mathbb{L},\pi_{\mathbb{L}})} - 2 \underbrace{( (2N)^2 + 2N^2-1)}_{\text{first-class}} -   \underbrace{2 ( (2N)^2-1)}_{\text{second-class}} = 2{\times} 2N^2\,.
\end{align}
Since this counting is in phase space, the number of degrees of freedom in configuration space is $2N^2$ which agrees with the dimension of the orbit momentum: $(2N)^2-1 - ( 2N^2-1) = 2N^2$ studied in section~\ref{sec:orbits}.

\medskip

Let us also compare the Hamiltonian action~\eqref{eq:canHmat} that contains the matrix Lagrange multiplier to the geometric action~\eqref{eq:geoL}.

The equations of motion following from the Hamiltonian action~\eqref{eq:canHmat} written in matrix form are
\begin{align}
    \dot{\mathbb{P}} &=0\,,\\
    \label{eq:xdot}
    \dot{\X} + \mathbb{P} \mathbb{L} + \mathbb{L}\mathbb{P} &=0\,,\\
    \label{eq:Hamc}
    \mathbb{P}^2 + m^2 \mathbb{I} &= 0\,.
\end{align}
The derivative here is with respect to the Lagrangian time variable.
The second equation~\eqref{eq:xdot} is uniquely solvable for $\mathbb{P}$ for generic $\mathbb{L}$; a closed form expression is not available but a formal series expansion can be obtained. Instead of solving for $\mathbb{P}$ we verify that the equations of motion agree with those of the geometric action~\eqref{eq:geoL}. We  deduce from~\eqref{eq:xdot}
\begin{align}
\label{eq:transPr}
    \dot{\X} + \frac{1}{m^2} \mathbb{P} \dot{\X} \mathbb{P} =0\,,
\end{align}
and since $\mathbb{P}$ is constant, the same equation also holds for $\ddot{\X}$
\begin{align}
    \ddot{\X} + \frac{1}{m^2} \mathbb{P} \ddot{\X} \mathbb{P} =0\,.
\end{align}
This equation can be understood as the projection of $\ddot{\X}$ to what in the uncoloured case would be the transversal directions to $\mathbb{P}$. If we can also argue that the `longitudinal' components
\begin{align}
\label{eq:long}
    \ddot{\X} - \frac{1}{m^2} \mathbb{P} \ddot{\X} \mathbb{P} 
\end{align}
vanish, then the Hamiltonian equations would imply $\ddot{\X}=0$ in agreement with the equations of the geometric Lagrangian~\eqref{eq:geoL} (in proper time gauge). In the uncoloured case the longitudinal combinations are gauge-variant and can be set to zero by a choice of gauge and for this reason we now study the gauge symmetries of the system~\eqref{eq:canHmat}. 

The gauge transformations entailed by the matrix constraint~\eqref{eq:Hamc} are for matrix parameter $\ci \mathbb{E}\in \mf{u}(N,N)$
\begin{align}
    \delta_\mathbb{E} \mathbb{P} =0 \,,\quad
    \delta_\mathbb{E} \X = \left\{ \mathbb{E}, \mathbb{P} \right\}\,,\quad
    \delta_\mathbb{E} \mathbb{L} = - \dot{\mathbb{E}}\,.
\end{align}
These gauge transformations are reducible since $\X$ is left invariant by any gauge parameter of the form $\mathbb{E}= \big[ \mathbb{F}, \mathbb{P}\big]$ for an arbitrary matrix $\mathbb{F}$ when~\eqref{eq:Hamc} is used. If this $\mathbb{F}$ is itself of the form $\big\{ \mathbb{G},\mathbb{P}\big\}$, it leads to a vanishing $\mathbb{E}$ when~\eqref{eq:Hamc} is used. This reflects the fact that the constraints form a reducible system of gauge transformations. Intuitively, the total effective number of gauge parameters is given by $(2N)^2\left[\sum_{n=0}^\infty (-1)^n\right]-1 = 2N^2-1$  by geometric sum regularisation and where the $-1$ is due to the identity component that has to be treated differently. The reducibility of the gauge transformations is tied to the reducibility of the constraint~\eqref{eq:Hamc} 

The longitudinal component~\eqref{eq:long} transforms as follows under gauge transformations
\begin{align}
    \delta_\mathbb{E} \left[ \ddot{\X} - \frac{1}{m^2} \mathbb{P} \ddot{\X} \mathbb{P} \right] = 2 \left\{ \ddot{\mathbb{E}},\mathbb{P}\right\}
\end{align}
and is clearly gauge-variant. Moreover, it is in the kernel of the `transversal' projection~\eqref{eq:transPr}.  We therefore expect to be able to gauge the longitudinal part to zero fully. For this to be correct, we need that any element in the kernel of the transversal projection is of the from $\{\ddot{\mathbb{E}},\mathbb{P}\}$ for some $\ddot{\mathbb{E}}$ with $\mathbb{P}$ satisfying~\eqref{eq:Hamc}. This can be checked to be true. 

We therefore conclude that the geometric action~\eqref{eq:geoL} and the Hamiltonian action~\eqref{eq:canHmat} with matrix constraint both imply the equation of motion 
\begin{align}
    \ddot{\X} =0
\end{align}
upon fully fixing the gauge invariances. In the geometric case this only requires the choice of proper time gauge $\frac1{2N}\Tr\big[\dot{\X}^2\big] = -1$ while for the Hamiltonian action this requires fixing all longitudinal components~\eqref{eq:long} to zero which is a larger set of gauge constraints. The two systems are therefore not equivalent and this agrees also with the analysis of which orbits they support that we carried out in section~\ref{sec:orbits}.

\subsection{\texorpdfstring{Examples for small $N$}{Examples for small N}}
\label{sec:smallN}

We shall now analyse the Lagrangian~\eqref{eq:L2n} for small values of $N$, beginning with the uncoloured $N=1$ case as a quick consistency check. It is useful to start with the component expression~\eqref{eq:L2comp}.

\subsubsection*{The uncoloured $N=1$ case}

For the case $N=1$, there are no $\mf{su}(N)$ generators $T^I$ and the symmetry group is just the usual Poincar\'e group in $2+1$ dimensions. Without $T^I$ generators there are no coordinates $x^a_I$, $y^I$ or associated momentum or Lagrange multiplier components. The Lagrangian~\eqref{eq:L2comp} reduces simply to 
\begin{align}
L = \dot{x}^a p_a + e( p^a p_a +m^2)  \,.
\end{align}
This agrees with the well-known free massive particle~\eqref{eq:LHamuc}. 
One can check similarly that the `orbit Lagrangian'~\eqref{eq:L1cexp1} reduces to~\eqref{eq:Lorbuc}. 

The only Lagrange multiplier is $e$ and the corresponding scalar constraint $p^a p_a +m^2=0$ has two branches of solution up to Lorentz rotation of the momentum, corresponding to positive and negative energy particles.

\subsubsection*{\texorpdfstring{The case $N=2$}{The case N=2}}

For the case $N=2$ we can rewrite~\eqref{eq:L2comp} by using $\mf{su}(2)\cong \mf{so}(3)$ to replace $\mf{su}(2)$ adjoint indices $I$ by $\mf{so}(3)$ vectors, e.g. $p_a^I \to \vec{p}_a$. Then the trace $\delta_{IJ}$ becomes the scalar product of vectors and $f^{IJK}$ becomes the cross-product. Note that $d^{IJK}=0$ for $N=2$. We will continue to write out the $\mf{so}(1,2)$ Lorentz indices that are raised and lowered with the Minkowski metric. 

Then~\eqref{eq:L2comp} becomes
\begin{align}
L &= \dot{x}^a p_a + \frac12 \dot{\vec{x}}^{\,a} \cdot \vec{p}_a -\frac12 \dot{\vec{y}}\cdot \vec{\pi} + e\left( p^a p_a +\frac12 \vec{p}^{\,a} \cdot\vec{p}_a -\frac12 \vec{\pi}\cdot\vec{\pi} + m^2\right) + \vec{\lambda}\cdot \vec{p}^{\,a} p_a  \nn\\
&\quad + f_a \,\vec{p}^{\,a} \cdot \vec{\pi} +\frac12 \vec{k}_a \cdot \left( 2\, \vec{\pi}\, p^a +\frac12 \varepsilon^{abc} \vec{p}_b\times \vec{p}_c\right)\,.\label{N2Lag}
\end{align}

The constraints implied by this Lagrangian from varying the Lagrange multipliers $e$, $\vec{\lambda}$, $f_a$ and $\vec{k}_a$ are 
\begin{subequations}
\begin{align}
\label{eq:N2a}
\text{(1 constraint)}&&p^a p_a +\frac12 \vec{p}^{\,a} \cdot\vec{p}_a -\frac12 \vec{\pi}\cdot\vec{\pi} + m^2 &=0 \,,\\
\label{eq:N2b}
\text{(3 constraints)}&&\vec{p}^{\,a} p_a &=0\,,\\[2mm]
\label{eq:N2c}
\text{(3 constraints)}&&\vec{p}^{\,a} \cdot \vec{\pi} &=0 \,,\\[2mm]
\label{eq:N2d}
\text{(3$\times$3 constraints)}&&4\vec{\pi}\, p^a + \varepsilon^{abc} \vec{p}_b\times \vec{p}_c &=0\,.
\end{align}
\label{eq:N2}
\end{subequations}
We have also written out the number of equations. 
The total of $16$ constraints is reducible. There are combinations that vanish identically without the use of the equations of motion and the simplest example is given by $\vec\pi\cdot
(\vec{p}^{\,a} p_a) -p_a(\vec{\pi}\, \vec{p}^{\,a})$ which is a single constraint. According to the general discussion in section~\ref{sec:orbits}, the solution space to the constraints is eight-dimensional for $N=2$ and we shall analyse this by solving them explicitly. 

We first consider the case when the particle has non-vanishing ordinary energy, so that $p^0\neq 0$. Then we can use all three equations of~\eqref{eq:N2b} to solve for $\vec{p}^{\,0}$ by
\begin{align}
\vec{p}^{\,0} = \frac{1}{p^0} \left( \vec{p}^{\,1} p_1 + \vec{p}^{\,2} p_2\right)\,.
\end{align} 

Similarly, the three $a=0$ components of~\eqref{eq:N2d} can be used to solve for $\vec{\pi}$ as
\begin{align}
\vec{\pi} = - \frac{1}{2p^0} \, \vec{p}_1\times \vec{p}_2 
\end{align}
Substituting these two solutions into~\eqref{eq:N2c} shows that they are satisfied automatically,
therefore the constraints are 1-stage reducible.
Moreover, substituting them into the $a=1$ and $a=2$ components of~\eqref{eq:N2d} imposes no further constraints either, reflecting again the reducibility.
 The only remaining constraint is~\eqref{eq:N2a} which yields the scalar constraint
\begin{align}
\label{eq:pquart}
p^a p_a +m^2+ \frac{1}{2p^0p_0}  \left( \vec{p}_1 p_1 + \vec{p}_2 p_2\right)^2  +\frac12 \vec{p}_1\cdot \vec{p}_1 +\frac12 \vec{p}_2\cdot \vec{p}_2 +\frac1{8p^0p_0}  \left(\vec{p}_1\times \vec{p}_2 \right)\cdot \left(\vec{p}_1\times \vec{p}_2 \right) =0\,.
\end{align}
This is one equation for the nine variables $p^a$, $\vec{p}_1$ and $\vec{p}_2$, leading in general to an eight-dimensional solution space. This can be checked explicitly for generic values by reading the above equation as  a quadratic equation for $p^0p_0$ and studying its discriminant. 
Among the four solutions of
$p_0=\pm\omega_{1}(p_1,p_2,\vec p_1, \vec p_2),\pm\omega_{2}(p_1,p_2,\vec p_1, \vec p_2)$ 
with 
\be 
    \omega_{1}(p_1,p_2,\vec p_1, \vec p_2)>
    p_1^2+p_2^2+\frac12\,\vec p_1^{\, 2}+
    \frac12\,\vec p_2^{\, 2} +m^2>\omega_{2}(p_1,p_2,\vec p_1, \vec p_2)>0\,.
\ee 
The two surfaces 
$p_0=\pm \omega_{1}(p_1,p_2,\vec p_1, \vec p_2)$ correspond to the ``momentum'' shells or orbits where
the energy is bounded from either below or above.
These orbits, 
denoted by $\mathcal O_{+2}$ and $\mathcal O_{-2}$, can be viewed as the colour extension of the positive and negative energy momentum shells.
The other two surfaces
$p_0=\pm \omega_2(p_1,p_2,\vec p_1, \vec p_2)$
rejoin
at the``equator" with $p_0=0$,\footnote{By
solving the system \eqref{eq:N2} 
with $p_0=0$, one
can obtain a seven dimensional surface corresponding to
this ``equator''.}
and form  a single orbit, denoted by $\mathcal O_0$.
The orbits $\mathcal O_{+2}, \mathcal O_0$ and $\mathcal O_{-2}$ can be
obtained from the representative momenta,
\begin{equation}
\label{orbrep}
    \widehat{\mathbb P}_{+2}=m\,
    \begin{pmatrix}
    \ci& 0 &0 &0\\
    0& \ci &0&0\\
    0&0&-\ci&0\\
    0&0&0&-\ci
    \end{pmatrix}\,,
    \quad 
       \widehat{\mathbb P}_{0}=m\,
    \begin{pmatrix}
    \ci& 0 &0 &0\\
    0& -\ci &0&0\\
    0&0&\ci&0\\
    0&0&0&-\ci
    \end{pmatrix}\,,
    \quad 
       \widehat{\mathbb P}_{-2}=m\,
    \begin{pmatrix}
    -\ci& 0 &0 &0\\
    0& -\ci &0&0\\
    0&0&\ci&0\\
    0&0&0&\ci
    \end{pmatrix}\,,
\end{equation}
and they correspond
to the homogeneous spaces,
\begin{equation}
    \mathcal{O}_{\pm2}\simeq 
    U(2,2)/(U(2)\times U(2))\,,
    \qquad 
    \mathcal{O}_{0}\simeq  
    U(2,2)/(U(1,1)\times U(1,1))\,.
\end{equation}
The first orbit $\widehat{\Pp}_{+2}$ is exactly the one analysed in~\eqref{eq:MPorb} for general $N$ and the representative momentum~\eqref{eq:rep1}. From~\eqref{eq:MPorb} one can see that $p^0$ is bounded from below. The representative $\widehat{\Pp}_{-2}$ of the third orbit  is the negative of $\widehat{\Pp}_{+2}$ and its orbit therefore has $p^0$ bounded from above. The middle orbit has points with $p^0=0$ and is a genuine new type of orbit for coloured Poincar\'e symmetry.

To recapitulate, the ``momentum'' space of the massive $N=2$ coloured particle
consists
of three distinct orbits of dimension eight, in agreement with the general discussion in section~\ref{sec:orbits}.

\subsection{Massless coloured particle}
\label{sec:massless}

We now consider the massless particle.
Again, we make the simplest choice for its representative `momentum'
assigning zero eigenvalues to all colour associated `momenta':
\be
	P_0=P_1=E\,, \qquad P_2=0\,, 
	\qquad P_a^I=0\,,
	\qquad Q^I=0\,,
	\label{massless}
\ee
where $E$ is the energy of this reference `momentum' state.
The little group is generated by $M_0+M_1, N_0^I+N_1^I$ and $T^I$,
hence, the remaining `broken' generators of the coloured Lorentz $\mf{su}(N,N)$ are
\be
	M_0-M_1\,,\qquad M_2, 
	\qquad M_0^I-M_1^I\,,\qquad M_2^I\,.
	\label{rem gen}
\ee
Again the eigenvalues \eqref{massless} define an element $\cK$ in the coadjoint space $\mf{su}(N,N)^*$:
\be
	\la P_a,\cK\ra=E\left(\delta_a^0+\delta_a^1\right),
	\qquad 
	\la P_a^I,\cK\ra=0\,,
	\qquad
	\la Q^I,\cK\ra=0\,.
\ee

Through the isomorphism \eqref{dual}, $\cK$ is mapped to 
\be
	H_\cK=E\left(P_0+P_1\right).
\ee
The massless orbit Lagrangian is given by
\be
\label{eq:massL}
	L_{\text{orb}}=\k\,\la\,\dot {\mathbb X}\,,\,
	{\rm Ad}^*_{b}\,\cK\ra
	=\frac{\k}{2N}\,\Tr\left[\dot {\mathbb X}\,b\,E\,(P_0+P_1)\,b^{-1}\right],
\ee
where $b$ is the exponentiation of a linear combination of the generators in \eqref{rem gen}:
\be
	b=\exp(\mathbb V)\,,
	\qquad
	\mathbb V=\begin{pmatrix} \mathbf V& -\mathbf V+\ci\,\mathbf W\\ \mathbf V+\ci\,\mathbf W  & -\mathbf V\end{pmatrix}\,.
\ee
Here, we packaged arbitrary linear  combinations of the generators in  \eqref{rem gen}
into $N\times N$ anti-hermitian matrices $\mathbf V$ and $\mathbf W$. The parameter $\kappa$ in~\eqref{eq:massL} is a dimensionful scale that has no further role at the classical level.

We first notice that
\be
	b\,E\,(P_0+P_1)\,b^{-1}
	=\mathbb R\,\mathbb R^\dagger\, \begin{pmatrix} \mathbf I & \mathbf 0\\ \mathbf 0 & -\mathbf I\end{pmatrix}\,,
\ee
where $\mathbb R$ is an $N\times 2N$ rectangular matrix,
\be
	\mathbb R=\sqrt{E}\,\exp(\mathbb V)\,\begin{pmatrix} \mathbf I \\ -\mathbf I \end{pmatrix}\,.
\ee
Since $\mathbb V$ can be decomposed as $\mathbb V=\mathbf V\otimes A+\ci\,\mathbf W\otimes B$ and
where
\be
	A= \begin{pmatrix} 1&-1 \\ 1&-1 \end{pmatrix},
	\qquad B= \begin{pmatrix} 0& 1\\ 1&0 \end{pmatrix},
\ee
satisfy
\be
	A^2=0\,,\qquad B^2=I\,,
	\qquad A\,B=-A\,,\qquad B\,A=A\,,
\ee
we find
\ba
    \exp(\mathbb V)\eq e^{\ci \mathbf W\otimes B}+
	\int_0^1 dt\,e^{\ci t\,\mathbf W\otimes B}\,(\mathbf V\otimes A)\,e^{\ci (1-t)\,\mathbf W\otimes B} \nn\\
	\eq	\cos(\mathbf W)\otimes I +\ci \sin(\mathbf W)\otimes B
	+\int_0^1 dt\,e^{\ci t\,\mathbf W\otimes I}\,(\mathbf V\otimes A)\,e^{-\ci (1-t)\,\mathbf W\otimes I}\,.
\ea
From the above,  the matrix $\mathbb R$ is obtained as
\be
	\mathbb R=\begin{pmatrix} \mathbf R+\ci \mathbf S\,\mathbf R \\  -\mathbf R+\ci \mathbf S\,\mathbf R \end{pmatrix}
	=\begin{pmatrix} \mathbf I+\ci \mathbf S \\ -(\mathbf I-\ci \mathbf S) \end{pmatrix}\,\mathbf R\,,
\ee
with two $N\times N$ hermitian matrices $\mathbf R$ and $\mathbf S$,
\be
	\mathbf R=\sqrt{E}\, e^{-\ci \mathbf W}\,,
	\qquad
	\mathbf S=-2\,\ci 
	\int_0^1 dt\,e^{\ci t\,\mathbf W}\,\mathbf V\,e^{\ci t\,\mathbf W}\,.
\ee
In terms of $\mathbf R$ and $\mathbf S$, the orbit is parametrised by 
\be
	\mathbb P=b\,E\,(P_0+P_1)\,b^{-1}
	=\begin{pmatrix} \mathbf I+\ci \mathbf S \\ -(\mathbf I-\ci \mathbf S) \end{pmatrix}\,\mathbf R^2\,
	\begin{pmatrix} \mathbf I-\ci \mathbf S & \mathbf I+\ci \mathbf S \end{pmatrix}\,.
	\label{R S para}
\ee
Note however that for a one-to-one correspondence between $\mathbf R$ and $\mathbf W$,
the former matrix should be invertible.
Foregoing the invertibility of $\mathbf R$, 
the parametrisation \eqref{R S para} contains other orbits besides the one generated by $E\,(P_0+P_1)$\,.

Now, let us look for algebraic constraints for $\mathbb P$, which lead
to the parametrisation  \eqref{R S para}.
From its explicit form of the parametrisation, one can verify
\be
\label{eq:P20}
	\mathbb P^2=0\,,
\ee
which is already obvious from its definition and the fact that $(P_0+P_1)^2=0$. 

A Lagrangian that contains all possible solutions of $\mathbb{P}^2=0$ is given by
\begin{align}
    L= \frac{1}{2N} \Tr\left[ \dot{\X}\,\mathbb{P} + \mathbb{L}\, \mathbb{P}^2 \right]\,,
\end{align}
which agrees with the Lagrangian~\eqref{eq:canHmat} restricted to $m^2=0$. This is completely analogous to what happens in the uncoloured case. 

Similarly
to the massive coloured particle,
the massless Lagrangian above 
contains several orbits,
but this time the structure is richer:
in the limit $m\to 0$, 
there appear new orbits of lower dimensions
on the boundary of the $N+1$ massive orbits of dimension $2N^2$.
More precisely, the massive orbits develop  
open regions in the massless limit
and the closure of these orbits contain
such sub-orbits.
The uncoloured analogue of this situation is as follows:
the massive momentum orbit, which is an upper hyperboloid,
becomes a cone in the massless limit.
Since this cone is open near the origin,
its closure contains the origin, which
is the new (trivial) orbit.

While in the uncoloured case the new sub-orbit is only
the trivial orbit, the situation is more interesting in the coloured case.
Here,  the sub-orbits are non-trivial
and they may even contain sub-sub-orbits. 
The sub-sub-orbits, in turn, may contain sub-sub-sub-orbits, etc, exhibiting an interesting inclusion structure that is familiar from the study of co-adjoint nilpotent orbits of $\mf{su}(N,N)$ that can be arranged in a Hasse (or closure) diagram~\cite{CollingwoodMcGovern}.
To quote the results directly,
the massless coloured particle
contains $k+1$ orbits of dimension $4Nk-2k^2$ where
$k$ varies from $0$ to $N$. The orbits with $k=N$ are the
massless limit of the massive orbits,
whereas the other orbits with $k<N$ are
newly appeared sub-orbits. The orbit with $k=0$
is the origin, corresponding to the zero-dimensional trivial orbit,
see also section~\ref{sec:orbits} for further discussions.

\subsection{Reductions to subspaces} 
\label{sec:subspaces}

The full coloured Minkowski space~\eqref{eq:ColMink} has coordinates $\X=(x^a, x^a_I, y_I)$ that transform under the coloured Poincar\'e algebra according to~\eqref{eq:CM}. We now consider the restriction to subspaces of coloured Minkowski space. These will necessarily break some of the coloured Poincar\'e symmetries. 

The most extreme subspace to be considered is ordinary Minkowski space obtained by setting $x^a_I=y_I=0$. Inspection of~\eqref{eq:CM} shows that this breaks the coloured Poincar\'e symmetry to the usual Poincar\'e symmetry.

A more interesting restriction is obtained by only setting $x^a_I=0$. In this case we still have the ordinary Minkowski coordinates $x^a$ and the coloured coordinate $y_I$ that we would like to think of as the space of an internal degree of freedom of a particle. Setting $x_I^a=0$ breaks the coloured Poincar\'e algebra down to the usual Poincar\'e symmetry and only coloured transformations generated by $N^I$ and $Q^I$, where we recall that the $N^I$ form an $\mf{su}(N)$ algebra. With only these generations, the equations~\eqref{eq:CM} reduce to
\begin{align}
    \delta x^a &= \varepsilon_{bc}{}^a \omega^b x^c + \alpha^a\,,\nn\\
    \delta y_I &= -f^{JK}{}_I \sigma_J y_K + \beta_I\,,
\end{align}
so that the two coordinates decouple with $x^a$ transforming under $\mf{iso}(2,1)$ and $y_I$ transforming under $\mf{su}(N) \inplus_{\text{adj}} \mf{su}(N)$.

Let us also consider what happens for the equations of motion~\eqref{eq:eoms} under the assumption that $x^a_I=0$ along with a vanishing of the conjugate momentum $p_a^I=0$. We keep, however, the Euler--Lagrange equation~\eqref{eq:kaIeq} for $k_{a I}$. The equations of motion simplify considerably in this reduction. We begin with~\eqref{eq:kaIeq} that becomes simply
\begin{align}
    p^a \pi^I =0
\end{align}
and there are two branches of solutions: $(i)$ with $\pi^I=0$ and $(ii)$ with $p^a=0$. Since we wish to describe a massive particle in ordinary Minkowski space, we focus on case $(i)$ where $p^a\neq 0$. Then all the remaining equations imply $\lambda^I=0$
\begin{subequations}
\begin{align}
    \dot{x}^a &= -2 e p^a\,,\\
    \dot{y}^I &= 2 k_{aI} p^a\,,\\
    p^a p_a + m^2 &=0 \,.
\end{align}
\end{subequations}
We see that the first equation is the usual propagation of a particle in Minkowski space and the last equation puts it on the standard mass-shell. The middle equation determines the propagation of the colour coordinate $y^I$ and is a bit reminiscent of the Wong equation~\cite{Wong:1970fu,Barducci:1976xq} that was developed to describe the coupling of a coloured particle to a Yang--Mills background. However, the equation contains the Lagrange multiplier $k_{a I}$ that should not be interpreted as the background field but takes the value zero. Therefore the colour component $y^I$ propagates freely. This is analogous to the coupling of an ordinary charged particle to an electro-magnetic background: The presence of a non-trivial, constant background requires the generalisation of the Poincar\'e algebra to the so-called Maxwell algebra where $[P_a,P_b]=Z_{ab}$, where the non-vanishing commutator of translations corresponds to the electro-magnetic field~\cite{Bacry:1970ye,Schrader:1972zd}. Non-constant backgrounds can be incorporated by further generalising the algebra~\cite{Bonanos:2008ez,Gomis:2017cmt}. We will comment more on the possible extension of the coloured Poincar\'e algebra to have non-trivial commutators of coloured translations in the conclusions.

\section{Free coloured particle in AdS background}
\label{sec:AdSpart}

In this section, we study the AdS$_3$ analog of the previous construction
of the coloured particle in three-dimensional Minkowski space-time.
For that, we begin with the coloured AdS$_3$ isometry algebra,
\be
	\mf{su}(N,N)\oplus \mf{su}(N,N)\,,
\ee
which has been considered in \cite{Gwak:2015vfb,Gwak:2015jdo}, see also appendix~\ref{app:AdS}.
One can view this algebra as the colour extension of $\mf{so}(2,2)\simeq \mf{su}(1,1)\oplus \mf{su}(1,1)$,
and the flat space algebra \eqref{eq:sdnocent} is related to the above by a In\"on\"u--Wigner contraction.
To identify the corresponding particle action, we follow the method of non-linear realisation (or coadjoint action). We are putting the AdS radius $\ell=1$ in this section.
Let us first consider the uncoloured case.

\subsection{Uncoloured particle in AdS}

One can obtain an action for a particle in AdS$_d\cong SO(d-1,2)/SO(d-1,1)$ simply by parametrising an $SO(d-1,2)$ element as 
\be
	g=t\,b\,h\,, 
	\label{AdS g}
\ee
with an $SO(d-1)$ element $h$ and
\be
	 t=\exp(x^0\,P_0)\,\exp(x^i\,P_i)\,,
	\qquad b=\exp(v^j\,M_{0j})\,,
\ee
and following the method of non-linear realisation, see for example~\cite{Coleman1,Coleman2,Ogievetski,Gomis:2006xw}. The elements $b$ and $h$ together form an element of the Lorentz group $SO(d-1,1)$.
It leads to the particle action with the coordinate system,
\be
	ds^2=\cosh^2r\left(-dt^2+
	dr^2+\sinh r^2\,d\Omega^2_{d-2}\right) \,,
\ee
where $r= \sqrt{x^ix_i}$ and $t=x^0$.
This metric does describe AdS$_d$ space-time, but the Lorentz symmetry $\mf{so}(d-1,1)$ is not manifest (or linearly realised).
Since the colour extension is based on the extension of the 3d Lorentz $\mf{so}(1,2)\simeq \mf{su}(1,1)$
to $\mf{su}(N,N)$, this choice of group element is not convenient
and we need to find another parametrisation with manifest Lorentz symmetry.

For Lorentz covariance, we can replace $t=\exp(x^0\,P_0)\,\exp(x^i\,P_i)$ in \eqref{AdS g}
by $t=\exp(x^a\,P_a)$.
Or more generally, we can consider\footnote{Note however
that the Lie group element $t$
depends not only on the function $f$ but in general also on the representation of $P_a$. Here, we take 
the defining representation.}
\be
	t=f(x^a\,P_a)\,.
\ee
In order that $t$ becomes a group element, we need to require
\be
	f(z)\,f(-z)=1\,.
	\label{f}
\ee
Any real function $f(z)$ with \eqref{f} would result in good AdS$_d$ particle actions but with
different coordinate systems.
For instance, in three dimensions the choice $f(z)=e^z$ corresponding to $t=e^{x^a\,P_a}$
gives the Lagrangian\footnote{A similar construction
can be found e.g. in \cite{Gomis:2020wrv} 
where curvature corrections to flat space-time
were studied.}
\be
	L=\dot x^\mu\,e_\mu^a\,p_a+ e\,(p^2+m^2)\,,
\ee
with the dreibein,
\be
	e_\mu^a=\frac{\sinh x}{x}\,\delta_\mu^a
	+\left(1-\frac{\sinh x}{x}\right)\frac{x^a\,x_\mu}{x^2}\,.
\ee
Here, we used the notation, $x^2=x^a\,x_a$ and $x=\sqrt{x^2}$.
Another convenient choice which we shall adopt for the colour extension
 is of Cayley transform type $f(z)=\frac{1+2z}{1-2z}$, that is
\be
	t=\frac{1+2\,x\cdot P}{1-2\,x\cdot P}\,,
	\label{h P}
\ee
with $x\cdot P=x^a\,P_a$\,.
Using identities of Pauli matrices, one can find
\be
	t^{-1}\,\dot t=
	\frac{2}{(1-x^2)^2}
	\left[
	\left((1+x^2)\,\dot x^a-2\,x\cdot \dot x\,x^a\right) P_a
	-2\epsilon_{abc}\,x^a\,\dot x^b\,M^c\right],
\ee
and the corresponding particle action,
\ba
	S\eq \int dt\,\langle m\,{\cal P}^0,g^{-1}\dot g   \rangle
	=\int dt\,\langle {\rm Adj}^*_b(m\,{\cal P}^0),t^{-1}\dot t  \,\rangle
	\nn\\
	\eq
	\int dt\left(\sqrt{p_i^2+m^2}\,e^0_\mu(x)+ p_i\,e^i_\mu(x)\right)\dot x^\mu\,,
	\label{AdS particle}
\ea
with the dreibein of the form,
\be
	e^a_\mu=2\,\frac{(1+x^2)\,\delta^a_\mu-2\,x_\mu\,x^a}{(1-x^2)^2}\,.
	\label{dreibein}
\ee
Combining with the negative energy sector corresponding to the coadjoint element $-m\,{\cal P}^0$,
we can rewrite the above action with a Lagrange multiplier $e$ as 
\be
	S[x,p]=\int dt\left( \dot x^\mu\,e^a_\mu(x)\,p_a+e (p^2+m^2)\right),
\ee
analogously to the flat space case.
The dreibein \eqref{dreibein} gives the metric, 
\be
	ds^2=4\,\frac{(1+x^2)^2\,dx^2-4\,(x\cdot dx)^2}{(1-x^2)^4}\,,
	\label{AdS metric}
\ee
and one can check that it describes AdS$_3$:
with the redefinition,
\be
	y^\mu=\frac{2\,x^\mu}{1+x^2}\,,
\ee
we can bring it to a more standard AdS metric,
\be
	ds^2=\frac{dy^2}{(1-y^2)^2}\,.
\ee
The AdS boundary is located at $y^2=1$ or equivalently $x^2=1$\,.

\subsection{Coloured particle in AdS}

Let us colour decorate the action \eqref{AdS particle}.
For that, we decompose an element of $SU(N,N)\times SU(N,N)$ as
\be
	g=t\,l\,
\ee
where $l$ is an element of the diagonal $SU(N,N)$, that is the colour extension of Lorentz subgroup.
The other element $t$ belongs to the off-diagonal part which can be interpreted as the colour extension of translation,
and we use the parametrisation analogous to \eqref{h P}
\be
	t=\frac{1+\mathbb X}{1-\mathbb X}\,.
\ee
Here, $\mathbb X=x^a_\hI\,P_a^\hI+y_I\,Q^I$ is an $2N\times 2N$ matrix with the condition \eqref{2N rep}.
In principle, we can take any function $f(\mathbb X)$ with \eqref{f}, but
the expression of $\dot h$ becomes complicated 
or implicit for a generic $f(z)$.  Different choices of $f(\mathbb X)$ corresponds to a field redefinition of $\mathbb X$, so
we consider only the simple case above.

Choosing the same representative of the coadjoint orbit, $m\,{\cal P}^0_{ 0}$, as in flat space case,
we find
\be
	S= m\int dt\, \langle {\rm Ad}^*_{l}(m\, {\cal P}^0_{0}), 
	t^{-1}\dot t\rangle\,,
\ee
as the action for coloured particle in AdS$_3$\,.
Using the isomorphism between $\mf{su}(N,N)$ and $\mf{su}(N,N)^*$,
we can rewrite the above as
\be
	S= \frac{1}{2N}\int dt\,
	\Tr\left[l\,\widehat{\mathbb P}\,l^{-1}\,t^{-1}\dot t\right],
\ee
where each element is expressed in the fundamental representation of $SU(N,N)$
and  $\widehat{\mathbb P}$ is as given in \eqref{eq:MPref}.
The $2N\times 2N$ matrix 
\be
	\mathbb P=l\,\widehat{\mathbb P}\,l^{-1}\,,
\ee
satisfies the condition $\mathbb P^2+m^2\,\mathbb I=0$,
but the latter contains other orbits besides the one determined by $m\,{\cal P}^0_{0}$.
The issue of algebraically describing the orbit at the cost of including other ones is the same as in the flat space case.
In the end, we find 
the AdS$_3$ analog of the coloured action
\eqref{eq:canHmat}
as
\be
	S=\frac1{N}\int dt\,\Tr\left[\mathbb P\left(\frac1{1+\mathbb X}\,\dot{\mathbb X}\,\frac1{1-\mathbb X}
	\right)
	-\mathbb L\left(\mathbb P^2+m^2\,\mathbb I \right)\right],
\ee
where $\mathbb X,\mathbb P\in \mf{su}(N,N)$
and $\mathbb L\in \mf{u}(N,N)$.
By a simple field redefinition, the action above can be written as
\be
	S=\frac1{N}\int dt\,\Tr\left[\mathbb P\,\dot{\mathbb X}
	-\mathbb L\,\Big( \big[(1-\mathbb X)\mathbb P(1+\mathbb X)\big]^2+m^2\,\mathbb I \Big)\right],
\ee
employing the standard Liouville form.

\section{Conclusions}
\label{sec:concl}

In this paper, we have defined a coloured extension of the three-dimensional Poincar\'e algebra. This algebra, after removing its center, is isomorphic to $\mf{su}(N,N)\inplus_{\adj} \mf{su}(N,N)$, where the first $\mf{su}(N,N)$ is the colouring of the Lorentz symmetry that acts on the coloured translations in the adjoint representation. 

We have defined Poincar\'e gravity based on this algebra using a Chern--Simons formulation where an interesting interplay between the colour degrees of freedom and the ordinary gravity degrees of freedom is implied by the structure of the algebra (and choice of bilinear form).\footnote{For the coloured AdS$_3$ case, see~\cite{Gwak:2015vfb,Gwak:2015jdo}.}

The coloured Poincar\'e algebra can also be used to define coloured massive and massless free particles. There are different types of coloured particles depending on the choice of co-adjoint orbit of the coloured Poincar\'e algebra, generalising positive and negative energy particles. There are various different Lagrangians depending on how many orbits one wishes to describe at the same time. 
We have also considered the extension to particles in coloured AdS space based on the algebra originally constructed in~\cite{Gwak:2015vfb}.

There are several interesting avenues to explore in order to further investigate the structures introduced here. One question, raised  in the introduction, is that of relating our analysis to the classical description of coloured particles in terms of the Wong equation~\cite{Wong:1970fu,Barducci:1976xq} and we have taken some first steps in this direction in section~\ref{sec:subspaces}, where we considered subspaces of coloured Minkowski space. Even more interesting would be to consider a further extension of the coloured Poincar\'e algebra to a coloured Maxwell algebra. The uncoloured Maxwell algebra allows for the description of a charged particle in an electro-magnetic background~\cite{Bacry:1970ye,Schrader:1972zd,Bonanos:2008ez,Gomis:2017cmt} and we expect the coloured Maxwell algebra to be the right framework for (constant) Yang--Mills backgrounds.

{\allowdisplaybreaks
The coloured Maxwell algebra has further generators $Z^{\hK},Z_a^{\hK}$ on top of those of the Poincar\'e algebra.  The commutation relations of the coloured Maxwell 
algebra can be obtained via semi-group expansion (see footnote~\ref{fn:exp}) of the algebra $\mf{u}(N,N)$ with
semi-group $S_E^{(2)}=\{\lambda_0,\lambda_1,\lambda_2,\lambda_3\}$
\begin{subequations}
\label{eq:cMax}
\begin{align}
\lb M_a^{\hI}, M_b^{\hJ} \rb &= \frac{1}2 \,\varepsilon_{ab}{}^c \,\hd^{\hI\hJ}{}_{\hK} \,M_c^{\hK} - \frac14\, \eta_{ab}\, \hf^{\hI\hJ}{}_{\hK}\, N^{\hK}\,,\\
\lb M_a^{\hI}, N^{\hJ} \rb &= \hf^{\hI\hJ}{}_{\hK} \,M_a^{\hK}\,,
\hspace{10mm}
\lb N^{\hI}, N^{\hJ} \rb = \hf^{\hI\hJ}{}_{\hK}\, N^{\hK}\,,\\
\lb M_a^{\hI}, P_b^{\hJ} \rb &= \frac{1}2 \,\varepsilon_{ab}{}^c \hd^{\hI\hJ}{}_{\hK}\, P_c^{\hK} -
 \frac14 \,\eta_{ab}\, \hf^{\hI\hJ}{}_{\hK}\, Q^{\hK}\,,\\
\lb M_a^{\hI}, Q^{\hJ} \rb &=
\lb N^{\hI}, P_a^{\hJ} \rb=\hf^{\hI\hJ}{}_{\hK} \,P_a^{\hK}\,,\\
\lb N^{\hI}, Q^{\hJ} \rb &= \hf^{\hI\hJ}{}_{\hK} \,Q^{\hK}\,,\\
\lb P_a^{\hI}, P_b^{\hJ} \rb &=
\lb M_a^{\hI}, Z_b^{\hJ} \rb=
\frac{1}2 \,\varepsilon_{ab}{}^c \,\hd^{\hI\hJ}{}_{\hK} \,Z_c^{\hK} - \frac14\, \eta_{ab}\, \hf^{\hI\hJ}{}_{\hK}\, Z^{\hK}\,,\\
\lb P_a^{\hI}, Q^{\hJ} \rb &=
\lb M_a^{\hI}, Z^{\hJ} \rb=
\lb N^{\hI}, Z_a^{\hJ} \rb=
\hf^{\hI\hJ}{}_{\hK} \,Z_a^{\hK}\,,\\
\lb Q^{\hI}, Q^{\hJ} \rb &=
\lb N^{\hI}, Z^{\hJ} \rb=
\hf^{\hI\hJ}{}_{\hK}\, Z^{\hK}\,.\\[-2mm]
\nn
\end{align}
\end{subequations}
All generators here are anti-hermitian and the generators of the semi-group $\lambda_i$ are hermitian. This algebra can also be embedded in a free Lie algebra construction of the type considered in~\cite{Gomis:2017cmt}. The algebra~\eqref{eq:cMax} is a quotient of the semi-direct sum of the $\mf{u}(N,N)$ with the free Lie algebra generated by the translations $(P_a^{\hI}, Q^{\hI})$ that transform in the adjoint of $\mf{u}(N,N)$. One could similarly consider other quotients leading to generalisations of the $\mathcal{B}_\infty$ algebra studied in~\cite{Salgado:2014qqa,Gomis:2019fdh} or to half a Kac--Moody algebra~\cite{Gomis:2019fdh}.}

Besides describing particles one may wonder whether there are coloured extensions of the actions of extended objects such as strings or branes (see, e.g., \cite{Gomis:2006xw}). In the non-relativistic limit one might also have to change the algebra, see for example
\cite{Brugues:2004an}.

Our analysis was restricted to classical particle dynamics and it would be interesting to extend the models to (free or interacting) field theories. A first question is which of our particle actions to quantise. As the different actions in section~\ref{sec:particle} comprise different orbits, this will correspond to including different irreducible representations at the quantum level.\footnote{When an action contains several orbits the classical phase space is in general a Poisson rather than a symplectic manifold and it would be interesting to apply Kontsevich's deformation quantisation methods~\cite{Kontsevich:1997} in this context.} The  constraints~\eqref{trconstraint} and \eqref{eq:matconst} are quadratic in the canonical variables and will lead to quadratic wave equations on the space of wavefunctions $\Psi(\Pp)$. However, as is evident from the detailed analysis of the case with $SU(2)$ colour in section~\ref{sec:smallN}, the wave equation on the reduced phase space can be of higher order. After solving most of the matrix-valued constraints, we arrived at the final constraint~\eqref{eq:pquart} arising in the case $N=2$ for the massive particle. This scalar equation can be read as a quartic condition on the momenta and therefore should translate into a fourth order Klein--Gordon equation in canonical quantisation of this mass-shell constraint. We note the general constraint~\eqref{eq:matconst} is matrix-valued and therefore would correspond to a matrix-valued Klein--Gordon operator whose direct interpretation is as an orbit condition. This situation is similar to what happens in $Sp(2N)$-invariant (higher-spin) field theory \cite{Vasiliev:2001dc}
(see \cite{Sorokin:2017irs} for a review),
where the matrix-valued field-equation
can be interpreted as 
the condition 
of the minimal nilpotent orbit
of $Sp(2N)$
(see \cite{Joung:2014qya} for
 discussions 
about the minimal nilpotent orbits 
of classical Lie algebras
in the context of higher spin field theory).

If the coloured particle 
is properly quantised,
it will  correspond to a representation of the group
$\CPoin_3(N)$. This representation
can be constructed analogously to
the induced scalar representation of the ordinary Poincar\'e algebra:
the representation space 
is the space of square integrable functions
on the orbit 
with a proper $SU(N,N)$ invariant measure on it.
The invariant measure can be obtained by 
solving the matrix-valued ``momentum-shell constraint'' $\mathbb P^2+m^2\,\mathbb I=0$,
and hence the hermitian product can be defined as
\begin{equation}
    \langle \psi_1,\psi_2\rangle
    =\int d\mathbb P\,
    \delta(\mathbb P^2+m^2\,\mathbb I)\,
    \psi_1(\mathbb P)^*\,\psi_2(\mathbb P)\,,
\end{equation}
where $d\mathbb P$ is the $((2N)^2-1)$-dimensional
 measure on $\mf{su}(N,N)$\,,
and $\delta(\mathbb P^2+m^2\,\mathbb I)$ is 
the $(2N^2-1)$-dimensional delta distribution
supported on the momentum orbits that we have described above.
The action of an element of the coloured Poincar\'e group, $(\mathbb \Lambda,\mathbb A)\in SU(N,N)\ltimes SU(N,N)$\,,
is given by
\begin{equation}
    (U(\mathbb \Lambda,\mathbb A)\,\psi)(\mathbb P)=
	e^{\ci\,\tr(\mathbb A\,\mathbb \Lambda^{-1}\,\mathbb P\,\mathbb \Lambda)}\,\psi(\mathbb \Lambda^{-1}\,\mathbb P\,\mathbb \Lambda)\,,
\end{equation}
and the unitarity follows from the invariance of the measure and the delta distribution,
\be
	d(\mathbb \Lambda^{-1}\,\mathbb P\,\mathbb \Lambda)=d\mathbb P\,,
	\qquad 
	\delta(\mathbb \Lambda^{-1}(\mathbb P^2+m^2\,\mathbb I)\,\mathbb \Lambda)=
	\delta(\mathbb P^2+m^2\,\mathbb I)\,.
\ee
A similar construction can be done in the AdS$_3$ case.\footnote{More specifically, 
the colour symmetry algebra for AdS$_3$ is $\mf{su}(N,N)\oplus \mf{su}(N,N)$,
and  the unitary representation 
would be of the form $\pi\otimes \pi$ 
where $\pi$ is a unitary irreducible representation
of $\mf{su}(N,N)$ (we take the same representation for both sides of $\mf{su}(N,N)$ because
we are describing scalar particles).
To identify $\pi$,
we can rely on the GK dimension, which
is the same in both flat and AdS$_3$ space.
Since GK dimension is additive
under tensor product,
the GK dimension of $\pi$
should be $N^2$, i.e., half of the dimension $2N^2$
of the momentum orbit.
In the massless case,
we have seen that the constraint $\mathbb P^2=0$ contains
momentum orbits of dimensions $4Nk-2k^2$
with $k=1,\ldots, N$ (except for the trivial
orbit $k=0$).
Therefore, the representations $\pi_k$
corresponding to massless coloured 
particle in AdS$_3$
should have GK dimension $2Nk-k^2$\,.
Interestingly, such representations 
can be realised by using the reductive dual pair $(U(N,N), U(k))$,
where we take the trivial representation
for the $U(k)$\,. For a review of dual pairs see~\cite{Basile:2020gqi}.
Disregarding the diagonal $U(1)$ of $U(N,N)$,
the corresponding $SU(N,N)$ representation
is the generalised Verma module induced from 
the one dimensional representation 
of the maximal compact subgroup $U(1)\times SU(N)\times  SU(N)
\subset SU(N,N)$ with
the $\mf{u}(1)$ eigenvalue $k\,N$\,.
Turning back to the massive case, 
the  representation is again given by the dual pair $(U(N,N),U(N))$ where we 
take other one-dimensional representations with eigenvalue $n$
for $U(1)\subset U(N)$ while keeping the trivial representation for $SU(N)\subset U(N)$\,.
Then, the corresponding $SU(N,N)$ representation 
is the generalised
Verma module induced from the one dimensional representation 
of $U(1)\times SU(N)\times  SU(N)
\subset SU(N,N)$ with
the $\mf{u}(1)$ eigenvalue $n+N^2$\,.
In the flat limit  where the AdS radius $\ell$ tends to $\infty$,
we take the limit $n\to \infty$ while keeping $n/\ell$ fixed as 
a constant proportional to the mass $m$.}

We also note that we are here dealing with an extension of a space-time symmetry and may wonder how this is compatible with the Coleman--Mandula theorem. Although we have not explicitly constructed a field theory with coloured Poincar\'e invariance, we do not expect any contradiction to the theorem since the space-time has been vastly extended from Minkowski space-time to coloured Minkowski space. We consider this extension to be a bit similar in spirit to what happens in supersymmetry (or higher spin theory).

Finally, our construction was strongly based on three-dimensional space-time and one big challenge is to generalise the construction to arbitrary space-time dimensions. 
As discussed here and in \cite{Gwak:2015vfb,Gwak:2015jdo}, non-abelian colour decorations require associative extensions of the algebra of isometries to start with.
Such extensions usually involve generators in non-trivial representations of the isometry algebra (see, e.g., \cite{Joung:2015jza}). Exceptions, i.e. extensions of isometry algebras involving only generators in trivial representations (that is, corresponding to symmetries of vector gauge fields), are possible in two dimensions (as discussed in section \ref{sec:CP3}) and
three dimensions (as we studied in this paper) as well as in five dimensions: the AdS$_5$ isometry algebra $\mathfrak{so}(2,4)\simeq \mathfrak{su}(2,2)$ (which is the same as conformal algebra in four dimensions) can be extended by a single generator to $\mathfrak{u}(2,2)$ allowing for colour generalisation to $\mathfrak{u}(2M,2M)\cong \mf{u}(2,2) \otimes \mf{u}(M)$. 
Interpreting the latter
as the coloured AdS$_5$ isometry algebra,
the momentum orbits
that we have considered in this paper can be viewed as
the phase spaces of coloured particles in AdS$_5$.
It is intriguing to realise
that the $N=2$ coloured AdS$_3$ particle, having the phase 
space $\cO_{\pm2}\times \cO_{\pm2}$,
can be interpreted as 
a particle on two copies of AdS$_5$
since the phase of the latter
is simply one factor $\cO_{\pm2}$. Here, $\cO_{\pm 2}$ refer to the (momentum) orbits discussed around~\eqref{orbrep}.

We also note that our analysis was restricted to scalar particles and it would be interesting to consider particles with spin as well. Spin corresponds to a non-trivial representation of the stability group which is here also enlarged compared to the uncoloured case. For instance, in the case of the representative momentum~\eqref{eq:MPref}, the stability algebra was $\mf{u}(1)\oplus \mf{su}(N)\oplus \mf{su}(N)$ and taking a non-trivial representation of $\mf{u}(1)$ would be analogous to ordinary spin but one could also envisage more colourful versions of spin by taking non-trivial representations of the $\mf{su}(N)$ parts. Extensions to superalgebras could also be interesting to explore.

\subsection*{Acknowledgements}

We would like to thank 
Thomas Basile for useful discussions.
JG acknowledges the hospitality
at
Max-Planck-Institut f\"{u}r Gravitationsphysik (Albert-Einstein-Institut), Potsdam where this work was started. 
EJ and KM are grateful  to the Erwin Schr\"odinger International Institute for Mathematics and Physics  for the hospitality during the program on ``Higher Spins and Holography''.
The work of JG
 has been supported in part by MINECO FPA2016-76005-C2-1-P 
and PID2019-105614GB-C21 and 
from the State Agency for Research of the
Spanish Ministry of Science and Innovation through the Unit of Excellence
Maria de Maeztu 2020-203 award to the Institute of Cosmos Sciences
(CEX2019-000918-M). 
The work of EJ was supported by National Research Foundation (Korea) through the grant  NRF-2019R1F1A1044065. 
KM is supported by the
European Union's Horizon 2020 research and innovation programme under the Marie
Sk\l odowska-Curie grant number 844265.
Part of this work was carried out while KM was at Max-Planck Institut f\"ur Gravitationsphysik and Scuola Normale Superiore di Pisa. 

\appendix

\section{Derivation from embedding in AdS algebra}
\label{app:AdS}

The uplift of the Poincar\'e algebra to an associative algebra proceeds in two steps. First we embed the $(2+1)$-dimensional Poincar\'e algebra in the corresponding AdS algebra $\mf{so}(2,2)\cong \mf{sl}(2,\reals)\oplus \mf{sl}(2,\reals)$ that we then extend to the associative algebra $\mf{a}=\mf{u}(1,1)\oplus \mf{u}(1,1)$. The coloured Poincar\'e algebra in $D=3$ is then obtained in the contracting limit where the AdS radius is sent to infinity. 

To see this in more detail consider first the AdS algebra 
\begin{align}
\label{eq:AdS3}
\lb M_a, M_b \rb = \varepsilon_{ab}{}^c M_c \,,\quad\quad
\lb M_a, P_b \rb = \varepsilon_{ab}{}^c P_c \,,\quad\quad
\lb P_a, P_b \rb = \frac{1}{\ell^2} \varepsilon_{ab}{}^c M_c \,,
\end{align}
where $M_a=\tfrac12 \varepsilon_{abc} M^{bc}$ are the dual Lorentz generators, $P_a$ the translation generators and $\ell$ denotes the AdS radius. Indices are raised and lowered with the $(-++)$ Minkowski metric in $D=3$. To bring out the $\mf{sl}(2,\reals)\oplus \mf{sl}(2,\reals)$ form of this algebra one defines the standard combinations
\begin{align}
J_a = \frac12 \left(M_a + \ell\, P_a\right) \,,\quad\quad
\tilde{J}_a = \frac12 \left( M_a - \ell \, P_a\right) \,,
\end{align}
such that
\begin{align}
\lb J^a, J^b \rb =  \varepsilon_{ab}{}^c J_c\,,\quad\quad
\lb \tilde{J}^a, \tilde{J}^b \rb =  \varepsilon_{ab}{}^c \tilde{J}_c\,,\quad\quad
\lb J^a,\tilde{J}^b \rb =0\,,
\end{align}
making the two commuting $\mf{sl}(2,\reals)$ manifest.

The AdS algebra~\eqref{eq:AdS3} can also be written via a tensor product of $\mf{sl}(2,\reals)$ with an associative abelian algebra $\mf{a}_\ell$ on two generators $\mathcal{I}$ and $\mathcal{J}$ satisfying
\begin{align}
\label{eq:IJads}
\mathcal{I}^2 =\mathcal{I} \,,\quad\quad
\mathcal{J}^2 =\frac{1}{\ell^2}\mathcal{I} \,,\quad\quad
\mathcal{I} \mathcal{J} = \mathcal{J} \mathcal{I} =\mathcal{J}\,.
\end{align}
Since $\mf{a}_\ell$ is abelian, one can turn $\mf{sl}(2,\reals)\otimes \mf{a}_\ell$ into a Lie algebra and this Lie algebra is isomorphic to the AdS algebra.\footnote{For $\ell=1$, the algebra~\eqref{eq:IJads} is identical to the semi-group $S_M^{(1)}$ studied in~\cite{Izaurieta:2006zz}.}
As the final step one defines
\begin{align}
\mf{A}_\ell = \mf{u}(1,1) \otimes \mf{a}_\ell
\end{align}
as an associative algebra. Sending $\ell \to \infty$ for the flat space limit and tensoring with the associative algebra $\mf{u}(N)$ one obtains the coloured Poincar\'e algebra in $D=3$ dimensions shown in~\eqref{eq:cPoin}.

\section{Free massive Poincar\'e particle}
\label{app:FP}

Here, we give a short exposition of the Lagrangian of a free massive particle built on the standard Poincar\'e group using non-linear realisations, see e.g.~\cite{Gomis:2006xw}. For easier comparison with the body of the paper we do everything in $D=3$ space-time dimensions using the algebra~\eqref{eq:Poin3} and write the algebra $\mf{so}(2,1)\cong \mf{su}(1,1)$ as $(2\times2)$-matrices.

\subsubsection*{Massive particle}

We begin with the construction from non-linear realisations. For a massive particle in rest frame, the momentum eigenvalues take the form $p_a=(m,0,0)$ with $SO(2)\cong U(1)$ stabiliser in the Lorentz group generated by $M_0$. The local subgroup $H$ of the non-linear realisation is thus $SO(2)$ and we write group elements in the form
\begin{align}
\label{eq:gP}
g = e^{x^a P_a} b \, h\,,
\end{align}
with $b=e^{v^i M_i}$ ($i=1,2$) the boost by the broken generators in the Lorentz group. We choose the gauge $h=\mathbb{I}$ for the $G/H$ coset representative.

The Poincar\'e algebra valued Maurer--Cartan form for this gauge-fixed coset element is
\begin{align}
\Omega = g^{-1} dg = b^{-1} \left(dx^a P_a\right) b + b^{-1} db = \Omega_{(P)}^a P_a + \Omega^a_{(M)} M_a\,.
\end{align}
The components of the Maurer--Cartan form are by construction invariant under the global Poincar\'e group acting from the left on the group element~\eqref{eq:gP}.

If one wants to construct a Lagrangian that is also invariant under the right-action of the unbroken local $SO(2)$ subgroup of the Lorentz group, one can therefore choose any component of the Maurer--Cartan form that is invariant under $M_0$. In the present case, there are two such candidates: $\Omega_{(P)}^0$ and $\Omega_{(M)}^0$. The former is properly invariant while the latter transforms as a gauge field into a total derivative and thus is only quasi-invariant. This quasi-invariance is special to $D=3$ dimensions, in general dimensions there is no such singlet from the Lorentz algebra decomposed under spatial rotations.\footnote{The form $\Omega_{(M)}^0$ is a Wess--Zumino term in three space-time dimensions~\cite{Schonfeld:1980kb,Plyushchay:1990yf,Mezincescu:2010gb,Batlle:2014sca}.} 
 We shall therefore ignore $\Omega_{(M)}^0$ and take the Lagrangian to be given by the pull-back of the component $\Omega_{(P)}^0$ to the world-line of the particle:\footnote{Note that $\Omega_{(P)}^0$ is nothing but the time column of a Lorentz boost acting on a general momentum.}
\begin{align}
\label{eq:m3}
L \, dt= m \left[ \Omega_{(P)}^0 \right]^*\,.
\end{align}

We note that the component $\Omega_{(P)}^0$ can also be obtained by pairing the space of translations, generated by the $P_a$, with its dual. In other words, using as the dual basis $\mathcal{P}^a$ with pairing
\begin{align}
\langle P_a , \mathcal{P}^b \rangle = \delta_a^b\,,
\end{align}
one has
\begin{align}
\Omega_{(P)}^0 = \langle \Omega_{(P)} , \mathcal{P}^0 \rangle
= \langle b^{-1}\left( dx^a P_a\right) b , \mathcal{P}^0 \rangle
= \langle \Ad_{b^{-1}}\left( dx^a P_a\right)  , \mathcal{P}^0 \rangle
= \langle \left( dx^a P_a\right)  , \Ad_{b^{-1}}^*\mathcal{P}^0 \rangle\,,
\end{align}
where we have used $\Ad_{b^{-1}}$ to denote the `adjoint' action of the Lorentz group element $b^{-1}$ on its vector representation when viewing the Poincar\'e group as a semi-direct product. Similarly, $\Ad^*_{b^{-1}}$ is the co-adjoint action on the dual space. 

Writing $x^a P_a =\X$ we therefore have the equivalent form of the Lagrangian~\eqref{eq:m3}
\begin{align}
\label{eq:Lpair}
L = \langle  \Pp , \dot{\X} \rangle\,,
\end{align}
where $\Pp$ is any element in the Lorentz group orbit of $m\mathcal{P}^0$. Here, it is important that the orbit of $\mathcal{P}^0$ can be written as the group coset $SO(2,1)/SO(2)$ by the orbit stabiliser theorem. This orbit can be parametrised by some parameters $v^i$ that then appear algebraically in the Lagrangian: the boost parameters. The derivative in $\dot{\X}$ denotes the derivative with respect to the parameter of the world-line and so we have explicitly carried out the pull-back.

Other orbits of the action of the Lorentz group on the space of momenta can also be considered and the form~\eqref{eq:Lpair} is universal in all cases. If the orbit is of the form $SO(2,1)/H$ for some subgroup $H$ then the Lagrangian corresponds to the non-linear realisation of the Poincar\'e group with local subgroup $H$.

This can also be treated as the $N=1$ case of the matrix analysis in the main body of the paper.
We take as reference momentum
\begin{align}
\widehat{\Pp} = m \mathcal{P}^0 = m \begin{pmatrix} \ci & 0 \\ 0 & -\ci\end{pmatrix}\,.
\end{align}
 The generalised boost~\eqref{eq:genb} becomes
\begin{align}
\label{eq:boostmass1}
b = \exp \left[ \frac12 \begin{pmatrix} 0 & v \\ \bar{v} & 0 \end{pmatrix}\right] 
= \begin{pmatrix}  \cosh\frac{|v|}{2} & v \frac{ \sinh\frac{|v|}{2} }{|v|}\\
\bar{v} \frac{ \sinh\frac{|v|}{2} }{|v|} &  \cosh\frac{|v|}{2}
\end{pmatrix}
\end{align}
with $v \in \cx$. The orbit of the reference momentum is
\begin{align}
b \,\widehat{\Pp} \, b^{-1} =  m\ci \begin{pmatrix} \cosh|v| & - v \frac{\sinh|v|}{|v|} \\
\bar{v} \frac{\sinh|v|}{|v|}  & - \cosh |v|
\end{pmatrix}
=\begin{pmatrix}
\ci \sqrt{|p|^2+m^2} & p \\ \bar{p} & -\ci \sqrt{|p|^2+m^2}
\end{pmatrix}\,,
\end{align}
where we have introduced $p = -m \ci \frac{\sinh|v|}{|v|} v$ as a change of coordinates for the algebraically appearing boost parameters.

The velocity vector is the matrix
\begin{align}
\dot{\X} = \begin{pmatrix}
\text{i}\dot{x}^0 & \dot{z}\\
\dot{\bar{z}} & -\text{i} \dot{x}^0
\end{pmatrix}
\end{align}
where $z=x^1+\text{i} x^2$. With this the Lagrangian 
\begin{align}
\label{eq:act11}
L &= \frac12 \Tr \left( \dot{\X}\, b\, \widehat{\Pp}\, b^{-1}\right)= m \left[ \dot{x}^0 \cosh|v| + \Im(\dot{z}\bar{v}) \frac{\sinh|v|}{|v|}\right]\nn\\
&= - \dot{x}^0 \sqrt{|p|^2+m^2} + \Re (\dot{z}\bar{p})
\,.
\end{align}
The equation of motion obtained by varying with respect to the algebraic variable $p$ is
\begin{align}
\frac{\dot{x}^0}{\sqrt{|p|^2+m^2}} \bar{p} = \dot{\bar{z}}\,.
\end{align}
Solving this for $p$ and substituting back into the Lagrangian leads to the expected
\begin{align}
\label{eq:B13}
L = - m \sqrt{ \left(\dot{x}^0\right)^2 - |\dot{z}|^2}\,.
\end{align}

\medskip
Going back to the action~\eqref{eq:act11} we can also perform another analysis that is closer to the main body of the text. For this we vary the Lagrangian~\eqref{eq:act11} with respect to $x^0$. This leads to 
\begin{align}
\label{eq:p0}
\sqrt{|p|^2+m^2} = p^0\,,
\end{align}
where $p^0$ is a constant of integration while the spatial $p$ is still a dynamical variable. As $p^0$ sits at the right place in the momentum matrix we deduce that it transforms as the zero component of the contravariant momentum $p^a$ would. Enforcing the square of the above relation with a Lagrange multiplier $e$  leads to the Lagrangian
\begin{align}
L=  \dot{x}^a p_a + e \left( p^a p_a + m^2\right)\,.
\end{align}
This Lagrangian has manifest Poincar\'e invariance. In this Lagrangian, all components of $p_a$ are algebraic  variables and we can integrate them out to obtain
\begin{align}
L = -\frac{1}{4e} \dot{x}^a\dot{x}_a + e m^2 =  - m \sqrt{-\dot{x}^a\dot{x}_a} \,,
\end{align}
where in the second step we have also integrated out $e$ (with a choice of square root).
This Lagrangian agrees with~\eqref{eq:B13} above.

\subsubsection*{Massless case}

Let us also consider the case of a massless particle in this formulation. The representative momentum is now given by
\begin{align}
\widehat{\Pp} = \frac12 \begin{pmatrix}
\ci E & E\\
E & - \ci E
\end{pmatrix}\,,
\hspace{10mm} \tr \left(\widehat{\Pp}\,\widehat{\Pp}\right) = 0\,,
\end{align}
and it is stabilised by the one-parameter group generated by $M_0+M_1=\begin{pmatrix}\ci &1 \\ 1& -\ci\end{pmatrix}$.

The boost can still be chosen as~\eqref{eq:boostmass1} and an arbitrary element of the orbit of the reference momentum is
\begin{align}
b\,\widehat{\Pp}\,b^{-1} = 
\begin{pmatrix}
\ci \omega & e^{\ci\theta} \omega \\
e^{-\ci \theta} \omega & -\ci \omega
\end{pmatrix}
\end{align}
in a convenient parametrisation. The Lagrangian becomes
\begin{align}
\label{eq:Lm0}
L &= \frac12 \Tr \left( \dot{\X}\, b\, \widehat{\Pp} \,b^{-1}\right)
= - \omega \left(  \dot{x}^0 - \Re \left(e^{-i\theta} \dot{\bar{z}}\right)\right)
\end{align}
The equation of motion when varying $\theta$ is
\begin{align}
 e^{2\ci \theta} = \frac{\dot{z}}{ \dot{\bar{z}}}\,,
\end{align}
leading to
\begin{align}
\label{eq:massless1}
L &=-\omega \left(  \dot{x}^0 -  \sqrt{ |\dot{z}|^2}\right)\,.
\end{align}
This enforces correctly the constraint $\dot{x}^a \eta_{ab} \dot{x}^b =0$, but is not quite the usual form for a massless Poincar\'e particle. 

Let us also go the alternative route for the Lagrangian~\eqref{eq:Lm0}, meaning that we vary with respect to $x^0$ first. This forces the energy of the particle to be constant
\begin{align}
\omega = p^0
\end{align}
for some constant $p^0$. We enforce the square of this relation as a constraint
\begin{align}
L =  p_0 \dot{x}^0 + p_i \dot{x}^i +e (\omega^2 - p_0^2)
= p_a\dot{x}^a +e p_a p^a\,,
\end{align}
where we have used that $\omega e^{\ci\theta}=p^1+\ci p^2$ represents the spatial momentum. Integrating out the covariant $p^a$ leads to expected action
\begin{align}
\label{eq:massless2}
L= -\frac1{4e} \dot{x}^a \dot{x}_a\,.
\end{align}

Both Lagrangians~\eqref{eq:massless1} and~\eqref{eq:massless2} can be obtained from the original~\eqref{eq:Lm0} (modulo the fact that for~\eqref{eq:massless2} a square for anti-particles was introduced).
Let us study their equivalence in general.
The equations of motion implied by~\eqref{eq:massless1} are
\begin{subequations}
\begin{align}
\label{eq1}
\dot{x}^0 - |\dot{\bf x}| &=0\,,\\
\label{eq2}
\frac{d}{d\tau} \omega &=0 \,,\\
\label{eq3}
\frac{d}{d\tau} \left( \frac{\dot{\bf x}}{|\dot{\bf x}|}\right) &=0\,,
\end{align}
\end{subequations}
while from~\eqref{eq:massless2} we get
\begin{subequations}
\begin{align}
\label{eq4}
\dot{x}^a \dot{x}_a &=0\,,\\
\label{eq5}
\frac{d}{d\tau} \left(\frac{ \dot{x}^a}{e}\right) &=0\,.
\end{align}
\end{subequations}
We first observe that~\eqref{eq1} and~\eqref{eq4} are equivalent, up to the choice of square root. Let us set
\begin{align}
\omega = \frac{\dot{x}^0}{e}
\quad\Leftrightarrow\quad
e = \frac{\dot{x}^0}{\omega} \underset{\eqref{eq1}}{=} \frac{|\dot{\bf x}|}{\omega}
\end{align}
and this is conserved consistently by~\eqref{eq2} and the first component of~\eqref{eq5}. The spatial components of~\eqref{eq5} are
\begin{align}
\frac{d}{d\tau} \left(\frac{\dot{\bf x}}{e}\right) = \frac{d}{d\tau} \left( \omega \frac{\dot{\bf x}}{|\dot{\bf x}|}\right) = \omega  \frac{d}{d\tau} \left(\frac{\dot{\bf x}}{|\dot{\bf x}|}\right) =0 
\end{align}
consistently with~\eqref{eq3} when using~\eqref{eq2}.

\section{General remarks on particle actions}

In this appendix, we collect some general remarks on the different ways of writing particle actions based on symmetries.

We go back to the action of nonlinear realisation or the coadjoint orbit,
\be
	S=
	\int dt\,\langle \varphi, g^{-1} \frac{d}{dt} g\rangle
\ee
where $g\in G$ and $\varphi\in \mathfrak{g}^*$. 
The coadjoint action of $a\in G$ is defined as
\be
	\langle {\rm Ad}^*_a \,\varphi,T\rangle
	=\langle\varphi,{\rm Ad}_{a^{-1}}\,T\rangle\,, \qquad \forall \,T\in \mathfrak{g}\,.
\ee
For a given $\varphi$, let us denote the stabiliser subgroup by
$G_\varphi=\{a\in G\,|\,{\rm Ad}^*_a\,\varphi=\varphi\}$
and its Lie algebra by $\mathfrak{g}_\varphi=\{T\in \mathfrak{g}\,|\,
{\rm Ad}^*_T\,\varphi=0\}$\,.
The dimension of the coadjoint orbit is simply,
\be
	{\rm dim}\,{\cal O}_\varphi={\rm dim}\,\mathfrak{g}/\mathfrak{g}_\varphi={\rm dim}\,{\mathfrak{g}}-{\rm dim}\,{\mathfrak{g}_\varphi}\,.
\ee
As an example, consider $G$ as the $D$-dimensional Poincar\'e group $ISO(D{-}1,1)$
and $\varphi=m\,{\cal P}^0$ where ${\cal P}^0$ is the dual basis generator corresponding to $P_0$.
Then,\footnote{Note that this contrasts with
\be
	\mathfrak{poin}_{P_0}
	={\rm Span}\{M_{ij}, P_0,P_i\}\,,
	\qquad
	\mathfrak{poin}/\mathfrak{poin}_{P_0}
	\sim {\rm Span}\{M_{0i}\}\,.\nn
\ee}
\be
	\mathfrak{poin}_{m\,{\cal P}^0}
	={\rm Span}\{M_{ij}, P_0\}\,,
	\qquad
	\mathfrak{poin}/\mathfrak{poin}_{m\,P_0^*}
	\sim {\rm Span}\{M_{0i},P_i\}\,.
\ee
Hence, we find the orbit has $2\times (D-1)$ dimensions, i.e. 
it is a phase space for $(D-1)$-dimensional mechanical system
so can describe the motion of a massive particle in $D$-dimensional  space-time.
If we want to see this more explicitly, 
we can consider an element of $g\in G$ such that
\be
	g=a\,b\,,\qquad b\in G_\varphi\,,
\ee
then,
\be
	\langle \varphi, g^{-1}dg\rangle =\langle \varphi, a^{-1}\,da\rangle+\langle \varphi, b^{-1}\,db\rangle\,.
\ee
The second term is closed and can only contribute when $G_\varphi$ has a non-trivial topology.
At the classical level, we can ignore this term.
Therefore, we see that the action depends only on $G/G_{\varphi}\simeq {\cal O}_\varphi$.
In order to get an explicit expression of the Lagrangian for a given $G$ and $\varphi$, 
it is sufficient to take an appropriate parametrisation of $G/G_{\varphi}\simeq {\cal O}_\varphi$. 
For instance, in the Poincar\'e case with $\varphi=m\,{\cal P}^0$, we can use
\be
	g= \exp( 
	x^a\,P_a)\,\exp(
	v^{i}\,M_{0i})\,R\,, 
\ee
where $R$ is a rotation element in $SO(D{-}1)$, and we can drop it as it contributes to the action
only as a boundary term.
We can also remove  $e^{\
x^0\,P_0}$ using
\be
	\exp(
	v^{i}\,M_{0i})\,e^{\tau\,P_0}
	=\exp\left(
	\tau\left(\cosh v\,P_0+\tfrac{\sinh v}{v}\,v^i\,P_i\right)\right)\exp(
	v^{i}\,M_{0i})\,.
\ee
With a choice of $\tau\,\cosh v=x^0$\,, we find 
\be
	g=\exp\left(
	\tilde x^i\,P_i\right)\exp(
	v^{i}\,M_{0i})\,
	\exp\left(
	-\tfrac{x^0}{\cosh v}\,P_0\right) R\,,
	\label{g reorder}
\ee
with
\be
	\tilde x^i=x^i-x^0\,\tfrac{\tanh v}{v}\,v^i\,.
	\label{tilde x}
\ee
Discarding the last two factors in \eqref{g reorder} since they belong to the stabiliser, we obtain the action,
\ba
	S\eq m\int dt\,\langle {\cal P}^0, \exp(
	-v^{i}\,M_{0i})\,\dot {\tilde x}^i\,P_i\,\exp(
	v^{i}\,M_{0i})\rangle\nn\\
		\eq - m \int dt\,\frac{\sinh v}{v}\,v_i\,\dot {\tilde x}^i\,.
	\label{tilde}
\ea
If we redefine $v^i$ as
\be
	\tilde p_i=-m\,\frac{\sinh v}{v}\,v_i\,,
\ee
we end up with an action without Hamiltonian,
\be
	S=\int dt\,\tilde p_i\,\dot{\tilde x}^i\,.
	\label{symp}
\ee
If we had not done the reordering of factors in $g$, we would have obtained
\be
	S=m\int dt\left(\cosh v\,\dot x^0-\frac{\sinh v}{v}\,v_i\,\dot x^i\right).
	\label{untilde}
\ee
One can also derive the above form of the action by
plugging \eqref{tilde x} into \eqref{tilde} and performing an integration by part.
Since the solution of \eqref{tilde} is simply,
\be
	\tilde x^i=\text{const},\qquad v^i=\text{const}\,,
\ee
the relation \eqref{tilde x} provides the general solution of 
the ordinary relativistic massive particle system \eqref{untilde}.

The fact that we can bring the action 
\eqref{untilde} to the form \eqref{symp} is not surprising because
the starting point of the action was the pullback of the Liouville one-form on the orbit.
Locally, we can always choose Darboux coordinates, then
the action will take the form of \eqref{symp}.
From this perspective, the dynamics of the system  
is inherited from the coordinate choice for the orbit,
or equivalently the group element $g$.
In order to make certain symmetries of an orbit manifest it is useful to choose appropriate coordinates on the orbit. Moreover, global properties of the orbit have to be taken into account.

\section{From the coloured Hamiltonian to the Lagrangian action}
\label{app:HamLag}

In the uncoloured case,
we can obtain a quadratic action 
by integrating out the momenta $p_a$.
We can try the same for the coloured particle~\eqref{eq:canHmat} 
where the equation for momenta 
becomes matrix equations
 \begin{equation}
     \dot{\mathbb X}+\mathbb P\,\mathbb L
     +\mathbb L\,\mathbb P=0\,.
     \label{Ham to Lag eq}
 \end{equation}
In order to solve this equation,
we assume the invertibility of $\mathbb L$ and deform the equation
by introducing $\epsilon$ as
\begin{equation}
    \mathbb P=-\mathbb L^{-1}\dot{\mathbb X}-\epsilon\,\mathbb L^{-1}\,\mathbb P\,\mathbb L\,.
\end{equation}
Eventually, we will take $\epsilon\to 1$ limit, but
 in the intermediate step we will keep $\epsilon$ 
 which plays the role of expansion parameter
 and regulator.
The solution of the above as a power seires in $\epsilon$ is
\be
    \mathbb P=-
    \sum_{n=0}^\infty (-\epsilon)^n\, \mathbb L^{-n-1}\dot{\mathbb X}\,\mathbb L^{n}\,.
\ee 
The Lagrangian~\eqref{eq:canHmat} becomes
\begin{align}
    2N\,L=&
    -\sum_{n=0}^\infty (-\epsilon)^n\,
    \Tr(\mathbb L^{-n-1}\dot{\mathbb X}\,\mathbb L^{n}\,\dot{\mathbb X}) \nn  \\
    &
    +\sum_{m=0}^\infty\sum_{n=0}^\infty
    (-\epsilon)^{m+n}\,
    \Tr(\mathbb L^{-n+m}\dot{\mathbb X}\,\mathbb L^{n-m-1}\,\dot {\mathbb X})+m^2\,\Tr(\mathbb L)\,.
\end{align}
We can perform a resummation for the double sum 
in the second line as 
\begin{align}
    \sum_{m=0}^\infty\sum_{n=0}^\infty
    (-\epsilon)^{m+n}\,
    \Tr(\mathbb L^{-n+m}\dot{\mathbb X}\,\mathbb L^{n-m-1}\,\dot {\mathbb X})
    =\frac1{1+\epsilon}\sum_{n=0}^\infty (-\epsilon)^n\,
    \Tr(\mathbb L^{-n-1}\dot{\mathbb X}\,\mathbb L^{n}\,\dot{\mathbb X})\,.
    \label{Tr quad}
\end{align} 
Above, we split the double sum into two parts, $m\ge n$ and $n>m$,
and each parts gives the prefactors 
$\frac1{1-\epsilon^2}$ and $\frac{-\epsilon}{1-\epsilon^2}$
respectively,
hence it is important to keep $|\epsilon|<1$ for the convergence of these terms. 
Note however that the final result (the sum
of two prefactors) is regular in $\epsilon$\,.
Using the above solution, we find the quadratic action as
\begin{equation}
   2N\, L=
    -\frac{\epsilon}{1+\epsilon}\,\sum_{n=0}^\infty (-\epsilon)^n\,
    \Tr(\mathbb L^{-n-1}\dot{\mathbb X}\,\mathbb L^{n}\,\dot{\mathbb X})+m^2\,\Tr(\mathbb L)\,,
\end{equation} 
As mentioned above,
the above series do not exhibit
any apparent divergence in the $\epsilon\to 1$ limit,
but we do not know whether the series converges.

For a more rigorous treatment, let us assume that the matrix $\mathbb L$ can be diagonalised as
\begin{equation}
    \mathbb L=\mathbb S\,{\rm diag}(\lambda_1,\ldots, \lambda_{2N})\,\mathbb S^{-1}\,,
    \qquad 
    |\lambda_1|>|\lambda_2|>\cdots>|\lambda_{2N}|>0\,.
\end{equation} 
Then $\mathbb P$ can be solved as
\begin{equation}
     (\mathbb S^{-1}\,\mathbb P\,\mathbb S)_{ij}
    =-\frac1{\lambda_i+\lambda_j}\,
    (\mathbb S^{-1}\,\dot{\mathbb X}\,\mathbb S)_{ij}\,,
\end{equation} 
and the Lagrangian reads
\begin{equation}
 2N\,L
 =-\frac12\sum_{i,j=1}^{2N}\frac{1}{\lambda_{i}+\lambda_{j}}\,
     (\mathbb S^{-1}\,\dot{\mathbb X}\,\mathbb S)_{ij}\,
     (\mathbb S^{-1}\,\dot{\mathbb X}\,\mathbb S)_{ji}
     +m^2\,\sum_{i=1}^{2N}\lambda_i\,.
\end{equation} 
Since the factor $(\lambda_i+\lambda_j)^{-1}$
prevents us to re-express the above as a trace,
we can try to expand it so that each term in the expansion
could be written as a trace.
However, for the convergence of the expansion
 we need to distinguish cases $i>j$, $i<j$ and $i=j$,
 and we can obtain at best
\begin{equation}
 2N\,L=-
 \sum_{n=0}^\infty (-1)^n\,
    \sum_{i>j}
    \lambda_i^{-n-1}\,
     (\mathbb S^{-1}\,\dot{\mathbb X}\,\mathbb S)_{ij}\,
     \lambda_j^{n}
     (\mathbb S^{-1}\,\dot{\mathbb X}\,\mathbb S)_{ji}
     -\frac12\,\sum_{i}\,\lambda_i^{-1}\, (\mathbb S^{-1}\,\dot{\mathbb X}\,\mathbb S)_{ii}^2
     +m^2\,\sum_{i=1}^{2N}\lambda_i\,.
\end{equation} 
If there were no restriction $i>j$, we could express the first series as
\be 
   -
 \sum_{n=0}^\infty (-1)^n\,
    \sum_{i,j}
    \lambda_i^{-n-1}\,
     (\mathbb S^{-1}\,\dot{\mathbb X}\,\mathbb S)_{ij}\,
     \lambda_j^{n}
     (\mathbb S^{-1}\,\dot{\mathbb X}\,\mathbb S)_{ji}
     =-
     \sum_{n=0}^\infty (-1)^n\, 
     \Tr(\mathbb L^{-n-1}\dot{\mathbb X}\,\mathbb L^{n}\,\dot{\mathbb X})\,,
\ee 
but because of this restriction we cannot.
Since the restricted sum $\sum_{i>j}$ can be understood as 
``one half'' of the unrestricted sum
$\sum_{i,j}$, one may have
an intuitive understanding of the expression
\eqref{Tr quad}.

Now, let us integrate out the Lagrange multiplier.
We can do it again in the diagonalised case. 
The equation for the eigenvalue $\lambda_i$ is
\begin{equation}
   \sum_{j=1}^{2N}\frac1{(\lambda_i+\lambda_j)^2}\, 
  A_{ij}
     +m^2=0\,,
     \qquad i=1,\ldots, 2N\,,
\end{equation}
where we defined 
\begin{equation}
    A_{ij}=A_{ji}= (\mathbb S^{-1}\,\dot{\mathbb X}\,\mathbb S)_{ij}\,
     (\mathbb S^{-1}\,\dot{\mathbb X}\,\mathbb S)_{ji}\,.
\end{equation}
For $N=1$, we find 
\begin{equation}
    \frac1{4\,\lambda_1^2}\,A_{11}
    +\frac1{(\lambda_1+\lambda_2)^2}\,A_{12}=-m^2\,,
    \qquad 
      \frac1{4\,\lambda_2^2}\,A_{22}
    +\frac1{(\lambda_1+\lambda_2)^2}\,A_{12}=-m^2\,.
\end{equation}
The equations can be re-expressed
as a quartic one so
they have four different solutions.
Since the Lagrangian is proportional
to the trace $\Tr(\mathbb L)=\lambda_1+\lambda_2$, we can focus on it and find
\begin{align}
   - 4\,m^2(\lambda_1+\lambda_2)^2
    &=4\,A_{12}+(\sqrt{A_{11}}\pm \sqrt{A_{22}})^2 \nn\\
    &=2(A_{11}+2\,A_{12}+A_{22})
    -(\sqrt{A_{11}}\mp \sqrt{A_{22}})^2
    \nn\\
    &=2\,\Tr(\dot{\mathbb X}^2)
    -
    \left((\mathbb S^{-1}\,\dot{\mathbb X}\,\mathbb S)_{11}
    \mp (\mathbb S^{-1}\,\dot{\mathbb X}\,\mathbb S)_{22}\right)^2\,.
\end{align}
When the relative sign in the last line
is postive
then it can be also expressed as a trace
(which is zero in this case),
and 
we find
\begin{equation}
    -4\,m^2(\lambda_1+\lambda_2)^2
    =2\,\Tr(\dot{\mathbb X}^2)
    -(\Tr(\dot{\mathbb X}))^2
    =2\,\Tr(\dot{\mathbb X}^2)\,.
\end{equation}
For $N=2$, we find a set of equations
equivalent to a single order-eight one, and hence would contain eight solutions.

\end{document}